\documentclass{aa}  
\usepackage{booktabs}

\usepackage{array}
\usepackage{natbib,twoopt}
\usepackage[breaklinks=true]{hyperref} 
\bibpunct{(}{)}{;}{a}{}{,}             
\makeatletter
  \newcommandtwoopt{\citeads}[3][][]{\href{http://adsabs.harvard.edu/abs/#3}%
    {\def\hyper@linkstart##1##2{}%
     \let\hyper@linkend\@empty\citealp[#1][#2]{#3}}}
  \newcommandtwoopt{\citepads}[3][][]{\href{http://adsabs.harvard.edu/abs/#3}%
    {\def\hyper@linkstart##1##2{}%
     \let\hyper@linkend\@empty\citep[#1][#2]{#3}}}
  \newcommandtwoopt{\citetads}[3][][]{\href{http://adsabs.harvard.edu/abs/#3}%
    {\def\hyper@linkstart##1##2{}%
     \let\hyper@linkend\@empty\citet[#1][#2]{#3}}}
  \newcommandtwoopt{\citeyearads}[3][][]%
    {\href{http://adsabs.harvard.edu/abs/#3}
    {\def\hyper@linkstart##1##2{}%
     \let\hyper@linkend\@empty\citeyear[#1][#2]{#3}}}
\makeatother
\usepackage{graphicx}
\usepackage{txfonts}
\usepackage{color}
\usepackage{multirow}
\usepackage{caption}
\usepackage{capt-of}

\begin{document} 

   \title{Toward a unified framework for helium observations and interpretation of atmospheric escape.} 
   \titlerunning{Comparing atmospheric escape He\,\textsc{i} observations with models}

   \author{Yann Carteret
          \inst{1} \&
          Vincent Bourrier\inst{1}
          } 

   \institute{Observatoire Astronomique de l’Université de Genève, Chemin Pegasi 51b, CH-1290 Versoix, Switzerland}

   \date{Accepted to A\&A}
 
  \abstract{Atmospheric escape is considered a key process in shaping exoplanet demographics and evolution. The near-infrared metastable helium triplet (He\,\textsc{i} at $\sim 10833$\,\AA) is one of the most powerful tracers of hydrodynamical outflows. 
  In recent years, a large variety of instruments have been used to detect and characterize expanding upper atmospheres. High-resolution (HR) spectrographs, which can spectrally resolve the lines of the triplet, have been mostly used to probe the dynamics of atmospheric escape. More recently, JWST (low-resolution, LR) detected extended outflows that were missed by previous ground-based observations due to night-length constraints. Since the observed transmission spectrum is computed from the ratio of stellar spectra degraded by instrumental convolution, we demonstrate that directly convolving a theoretical transmission spectrum with the instrument's profile leads to inference biases as resolution decreases. This conclusion is particularly important for atmospheric escape studies but also holds in the more general context of LR atmospheric retrievals. Following the proper comparison methodology, we then investigated the complementarity of HR and LR observations for atmospheric escape measurements, comparing the capability of three instruments (NIRPS, JWST/NIRSpec, and JWST/NIRISS) to detect and characterize outflows. We find that NIRPS and JWST/NIRISS are mostly sensitive to the same helium signatures, while JWST/NIRSpec improves the detection limit of excess absorption for faint targets. Studying different outflow configurations, we show that LR space-based measurements cannot be used to infer the dynamics of the upper atmosphere close to the planet, and are best employed to measure the spatial extent of outflowing tails. We thus highlight the complementarity between HR and LR observations of atmospheric escape, the latter improving baseline reconstruction.}


   \maketitle
%
\section{Introduction}

Since the discovery of the first exoplanet atmosphere \citep{Charbonneau2002}, the field has evolved toward in-depth characterization of molecular composition and dynamics of atmospheres. The extraction of atmospheric transmission spectra relies on the comparison between the in-transit (planet-contaminated) and out-of-transit (unadulterated) stellar spectra averaged outside of the planetary transit, referred to as the baseline. Observations are either performed at high spectral resolution (HR; $\mathcal{R}>20\,000$), only accessible from ground-based spectrographs (e.g. CARMENES \citealt{Quirrenbach2016}, ESPRESSO \citealt{Pepe2021}, NIRPS \citealt{Bouchy_2025}), or medium ($2\,000<\mathcal{R}<20\,000$) and low spectral resolution (LR; $\mathcal{R}<2\,000$), mostly achieved by space-based instruments (e.g. HST \citealt{HST}, JWST \citealt{jwst}). These two observing strategies are usually complementary, as they probe different layers of the atmosphere. Indeed, HR observations are sensitive to cores of atomic and ionic lines, tracing the uppermost layers, where molecules are mostly dissociated. Conversely, LR spectroscopy detects mostly molecular bands and the wings of these lines, formed at higher pressures, and thus deeper within the atmosphere. These broadband features are less accessible from the ground \citep[but still observable, see e.g.][]{Rackham2017,DiamondLowe2018} due to the loss of flux calibration in ground-based observations because of Earth's atmosphere and telluric contamination. Nevertheless, HR observations enable the detection of a broad range of species located in the upper atmosphere, from escaping light elements such as hydrogen and helium \citep{Lavie2017,Guilluy2023}, to heavier metals \citep{Hoeijmakers2018,Sing2019,Ehrenreich2020,Prinoth2023} and some molecules \citep{Snellen2010,Birkby2013,Hood2024}. Resolving the intrinsically narrow line cores of these species allows for a detailed study of atmospheric chemistry and dynamics \citep{Seidel2025}. On the other hand, LR observations enable the study of clouds, hazes, and molecular composition detected through broad spectral features \citep{Lecavelier2008,Sing2016,Helling2020}.

Due to Earth's atmospheric absorption, the first observations of atmospheric escape were only possible using space instruments through the Lyman-$\alpha$ line \citep{Vidal2003}. These measurements mainly probed the exosphere, where interactions between atoms are sparse.
The metastable helium triplet lines (He\,\textsc{i}, $\sim 10833$\,\AA, \citealt{Seager2000,Oklopcic2018}) later emerged as a powerful tracer of highly irradiated, expanding upper atmospheres \citep{Allart2018,nortmann2018}. Since He\,\textsc{i} decays to its ground state within a few hours, it traces the hydrodynamical regime of the extended atmosphere where it can form, most often associated to the thermosphere. Located in the near-infrared, the triplet is readily accessible from the ground, observed using HR spectrographs which can spectrally resolve the individual lines or ultra-narrow band photometry \citep{vissapragada2020}. More recently, the JWST has demonstrated that space-based LR observations of the triplet can not only yield new detections around small planets \citep{ahrer2025}, but also estimate the outflow dynamics \citep{Allart2025Nat}.

Despite being two distinct observing strategies, HR and LR observations of He\,\textsc{i} probe the same layers of the atmosphere and measurements can be compared. In this paper, we address the question of observing and interpreting atmospheric escape using the He\,\textsc{i}. With both HR and LR observations available and exploited by the community, it is essential to understand to which extent the two observing strategies are complementary, in order to optimize observing time and interpret them jointly. In particular, we compare the performance of LR and HR instruments in detecting the He\,\textsc{i}. Furthermore, we discuss how models should be compared to LR and HR measurements to be properly interpreted and avoid any biases. In this study, we use the Near InfraRed Planet Searcher \citep[NIRPS,][]{bouchy_near-infrared_2017,Bouchy_2025}, and instruments of the James Webb Space Telescope (JWST): JWST/NIRSpec (Near-Infrared Spectrograph) and JWST/NIRISS (Near-InfraRed Imager and Slitless Spectrograph) \citep{jwst}; as representative of, respectively, HR and LR instruments, considering they are among the latest-generation instruments currently operating in this band pass.

In Sect.~\ref{sec:method}, we describe the methodology employed to generate synthetic observations throughout this paper. In Sect.~\ref{sec:instrumental_convolution}, we focus on the biases induced by the current standard method to compute the absorption spectrum at both HR and LR. Sect.~\ref{sec:high_vs_low} compares the capabilities of HR and LR in detecting and characterizing atmospheric escape, in particular the geometry of the outflow and thermospheric properties. In Sect.~\ref{sec:baseline}, we study the choice of baseline for atmospheric escape observations. Finally, in Sect.~\ref{sec:conclusion} we summarize our findings and present our conclusions.

\section{Methodology}
\label{sec:method}
In this section we describe our procedure to generate synthetic flux spectra during planetary transits, using the \texttt{EvE} code \citep{Bourrier2013,Bourrier2015,Dethier2023}, and their associated error-bars. This procedure is common to all the sections of this paper. 

\subsection{Stellar spectra simulations}

The \texttt{EvE} code simulates an exoplanetary transit while accounting for the geometrical effects associated with its three dimensional (3D) nature. The star is described by a 2D regular square-cell grid, with each cell assigned a specific local spectrum. This approach accounts for inhomogeneities across the stellar surface and for the local occultation by the three different planetary regimes in \texttt{EvE}: the opaque layers which absorbs 100\% of the stellar flux, the extended thermosphere in fluid regime, and the escaping exosphere with particle prescription. Based on the spectral profile associated with each stellar cell, and the optical depth of the planetary atmosphere computed along their line of sight to the Earth, \texttt{EvE} generates a time series of disk-integrated spectra at both high spectral resolution and temporal cadence. The code then convolves these spectra with the chosen instrumental response and resamples them both temporally and spectrally, so that they can be compared with observations \citep[see][for more details]{Dethier2023,Carteret2024}.

In this study, we use as a reference the hot-Saturn WASP-69\,b, adopting its system parameters from \citet{Allart2025}. This bright system \citep[J $\sim$ 8;][]{Anderson2014} has been extensively characterized through numerous observations of He\,\textsc{i} at both low- \citep{vissapragada2020,levine2024} and high-resolution \citep{nortmann2018,allart2023,guilluy2024,masson2024,tyler2024}, with one of the highest signal-to-noise ratio (S/N) among known exoplanets in the He\,\textsc{i} \citep[$\sim65$ reported in][]{Allart2025}. Moreover, it was theoretically predicted to be the best system for atmospheric escape measurements in the He\,\textsc{i} band \citep{linssen2024}. For these reasons, it is an ideal system for assessing instrument capabilities and evaluating potential biases when comparing atmospheric escape observations with models.

The stellar grid used in this work is tiled based on observations from \citet{Allart2025}, accounting for the variations in local stellar spectra across the photosphere. Indeed, the disk-integrated spectrum is generally a poor proxy for local spectra, which vary along the transit chord and across the full stellar disk, resulting in planet-occulted line distortions \citep[POLDs][]{Dethier2023} in classical transmission spectra. In particular, we account for the following POLD sources in our stellar grid: the Rossiter-McLaughlin effect \citep[RM;][induced by stellar rotation]{Rossiter1924,McLaughlin1924}, center-to-limb
variations (CLVs) of the line profiles \citep[e.g.,][]{Vernazza1981}, and variations of the broadband continuum \citep[limb darkening, LD;][]{Knutson2007,Morello2017}. 

Thermospheric profiles are generated using the \texttt{p-winds} \citep{DosSantos2022} code, adopting arbitrary parameters that roughly reproduce the amplitude of the observed signal in previous studies. The H/He ratio is fixed at 90:10\footnote{While the shape of the metastable helium profile is affected by this assumption, the general conclusions drawn across the paper does not depend on this choice.}, and the exobase location depends on the thermosphere geometry (see Fig.~\ref{fig:shape}). The exosphere is described by a Monte Carlo particle simulation, initialized with the properties of the thermosphere at the exobase and shaped by the stellar XUV flux. Finally, the structure of the leading and trailing tails is assumed to be shaped to first order by gravity and approximated by tube-like structures. Their geometry in the planetary rest frame is determined by free particle motion, for which we solve the 2D gravitational equations. The tails are thus parametrized by an escape velocity and an angle in the orbital plane between the perpendicular to the orbit and the tail axis. 

\subsection{Mock observations}

We simulate time series of flux spectra $f_\mathrm{in}$ at high spectral and temporal resolution during a single transit. We convolve the time series of synthetic flux spectra with a Gaussian profile whose width is set by the resolving power $\mathcal{R}$ of the instrument. We consider in this study NIRPS and JWST, which are respectively representative of high-resolution (HR) ground-based and low-resolution (LR) space-based approaches to He\,\textsc{i} observations. For HR observations, we produce synthetic spectra as observed by NIRPS in HE mode ($\mathcal{R}\sim 75\,000$; pixel size $\Delta\lambda\sim 0.036\,\mathrm{\AA}$). For LR we differentiate between JWST/NIRSpec in the BOTS (Bright Object Time Series) mode, G140H grating, and F070LP filter (hereafter NIRSpec, $\mathcal{R}\sim 2\,050$; $\Delta\lambda\sim2.39\,\mathrm{\AA}$) and JWST/NIRISS in SOSS (Single Object Slitless Spectroscopy) mode (hereafter NIRISS, $\mathcal{R}\sim 650$; $\Delta\lambda\sim9.38\,\mathrm{\AA}$ near the helium triplet). The convolved time series is then resampled to the instrument's wavelength grid. Finally, the absorption time series $\mathcal{A}$ is computed as the ratio of the in-transit spectral time series to the unocculted stellar spectrum $f_\star$, processed identically:
\begin{equation}
\label{eq:abs_sp}
    \mathcal{A}(\lambda,t) = 1-\frac{R_s\big(G_\mathcal{R}\ast f_\mathrm{in}(\lambda,t)\big)}{R_s\big(G_\mathcal{R}\ast f_\star(\lambda)\big)},
\end{equation}
where $R_s$ is the resampling function and $G_\mathcal{R}$ the Gaussian kernel.

Throughout this paper, we work with a time-averaged absorption spectrum and a He\,\textsc{i} light curve. The averaged absorption spectrum is obtained by taking the temporal mean of the absorption time series between second and third contact, when the geometric planetary occultation is consistently full. To properly isolate the atmospheric absorption, spectra are corrected for LD before averaging as detailed in \citet{Mounzer2022} and \citet{Dethier2023}. For NIRPS, prior to the computation of the average absorption spectrum, we resampled the flux time series into 5-minutes windows in order to match typical observational strategy.

On the other hand, the light curves are computed by integrating the flux over three wavelength ranges specific to each instrument (adjusted to the line spread function), reported in Table~\ref{tab:error}. We average the flux time series within 12-minutes bins for NIRPS (typical windows used for observations), while JWST is fixed to 5 minutes. This is motivated by the higher sensitivity of JWST compared to NIRPS, which ultimately allows for a better temporal sampling. We then integrate spectrally the excess absorption
\begin{equation}
    \mathrm{lc}_{\mathrm{He}}(t) = \frac{\int \mathcal{A}(\lambda,t) - (1-\mathrm{lc}_{\mathrm{white}}(t))\, \mathrm{d}\lambda}{\int \mathrm{d}\lambda},
\end{equation}
with $\mathrm{lc}_{\mathrm{white}}$ the broadband white light curve, as this traces more accurately the atmospheric contribution. When the integration is performed directly on the average excess absorption spectrum, we refer to it as the equivalent width (EW). The three spectral ranges used in this study correspond to the core, and the blue/red wings of the helium lines. At LR, we expect the absorption signal in the wing ranges to originate from convolution broadening, whereas at HR it arises mainly from the projected dynamics of the outflow along the transit chord. The core range was thus selected based on previous observations to maximize the integrated helium signal.

The set of mock observables is finally obtained by computing uncertainties on the averaged spectra and light curves, propagating errors from the flux time series. Error-bars are computed for one transit of WASP-69\,b, and then scaled to the J magnitude using the S/N per second and the instrument exposure time $t_\mathrm{exp}$ as
\begin{equation}
    \sigma_f(J) = \sigma_{f,W69}
    \frac{\mathrm{S/N}_{W69}}{\mathrm{S/N}_{1s}(J)\sqrt{t_\mathrm{exp}}}.
\end{equation}
For NIRPS, we use the S/N and error-bars from \citet{Allart2025} as reference, provided for each exposure of the three observed transits of WASP-69\,b. The errors for different J magnitudes are then estimated using the observational data from NIRPS GTO \citep{Bouchy_2025} to calibrate the S/N as a function of magnitude. These uncertainties directly reflect the average observational conditions, which is more realistic than estimates from the Exposure Time Calculator (ETC) for ground-based facilities (due to varying weather conditions). For both JWST instruments, we use \texttt{Pandexo}\footnote{Available online at \href{https://exoctk.stsci.edu/pandexo/calculation/new}{Pandexo}.} \citep{pandexo} to calculate the errors on the flux spectrum of WASP-69\,b for one transit under the recommended observational strategy and the ETC\footnote{Available online at \href{https://jwst.etc.stsci.edu/}{JWST ETC}} to compute the S/N as a function of the stellar J magnitude.

\begin{table}
\small
\caption{Parameters associated with the three instruments of this study.}
\centering{
\begin{tabular}{c|c|c|c}
         & NIRPS  & NIRSpec/G140H & NIRISS/SOSS \\ \hline
$\mathcal{R}$ & 75\,000 & 2\,050 & 650 \\
$\Delta\lambda$ [\AA]     & 0.036  & 2.39 & 9.38 \\
\hline
$t_\mathrm{exp}$ [s]      & 720 & 300 & 300  \\
$\lambda_b$ [\AA]     & [-2.03,-0.53]  & [-3.31,-0.92] & [-10.75,-1.37] \\
$\lambda_c$ [\AA]     & [-0.53,0.97]  & [-0.92,1.47] & [-1.37,8.00] \\
$\lambda_r$ [\AA]     & [0.97,2.47]  & [1.47,3.86] & [8.00,17.38] \\
$\Delta\lambda_\mathrm{lc}$ [\AA] & 1.5 & 2.39 & 9.38
\end{tabular}}
\tablefoot{$\mathcal{R}$ is the resolving power, $\Delta\lambda$ is the pixel size, $t_\mathrm{exp}$ is the exposure time of the simulated light curves, $\lambda_{b/c/r}$ is the spectral integration range (centered on 10\,833\,\AA) of the blue wing / core / red wing light curves, and $\Delta\lambda_\mathrm{lc}$ is the spectral integration size of the light curves.}
\label{tab:error}
\end{table}

\section{Absorption spectrum and instrumental convolution}
\label{sec:instrumental_convolution}
In this section, we explore the impact of the geometrical model assumed for the planetary orbit and atmosphere on the computed absorption spectrum. In particular, we compare the results given by a 1D prescription, commonly adopted in the literature, with a more advanced treatment including a 3D description of the orbit. We also assess the influence of the spectral convolution on the resulting absorption spectrum.

\subsection{Thermospheric parameters}
\label{sec:convolution_thermosphere}
We generated synthetic observations at HR (NIRPS) and LR (NIRSpec) with \texttt{EvE}, as described in Sect.~\ref{sec:method}. The thermosphere is generated with \texttt{p-winds} at a temperature of 9750\,K and a mass-loss rate of $10^{12}$\,g/s, roughly reproducing previous observations \citep{Allart2025}, and injected into \texttt{EvE} without including the exosphere. As a reminder, our modeling naturally includes RM, CLVs, and LD effects in these synthetic observations, as well as a 3D description of the planetary orbit and thermosphere.

From the synthetic average absorption spectrum, we generate a mock observation. Let us consider the number of photons $N_\gamma$, which is related to the spectrum in flux units by $N_\gamma/\sigma_{N_\gamma} = S/N = f/\sigma_f$. We estimate the errros on the number of photons collected as $\sigma_{N_\gamma}= \sqrt{N_\gamma}$, with $f$ and $\sigma_f$ known from \texttt{EvE}. Using the number of photons computed from the \texttt{EvE} averaged flux spectrum, we draw a random realization from a Poisson distribution. Once converted back to flux values, this yields a new averaged absorption spectrum (and its associated error), randomly and conservatively reconstructed within the initial uncertainties. This represents one mock observation as seen by a given instrument under realistic observational conditions.

We then generated a grid of \texttt{p-winds} models using its built-in the radiative transfer. The grid of models spans a wide parameter space in temperature and mass-loss, with the H/He ratio and thermospheric radius fixed for simplicity (see Sect.~\ref{sec:method}). Since \texttt{p-winds} (and most atmospheric codes) does not incorporate the stellar flux in the computation of the averaged absorption spectrum, we directly convolve and resample it (as commonly done in the literature), unlike the more accurate approach we advocate in Eq.~\ref{eq:abs_sp}.

\begin{figure*}[ht]
    \includegraphics[width=1.75\columnwidth]{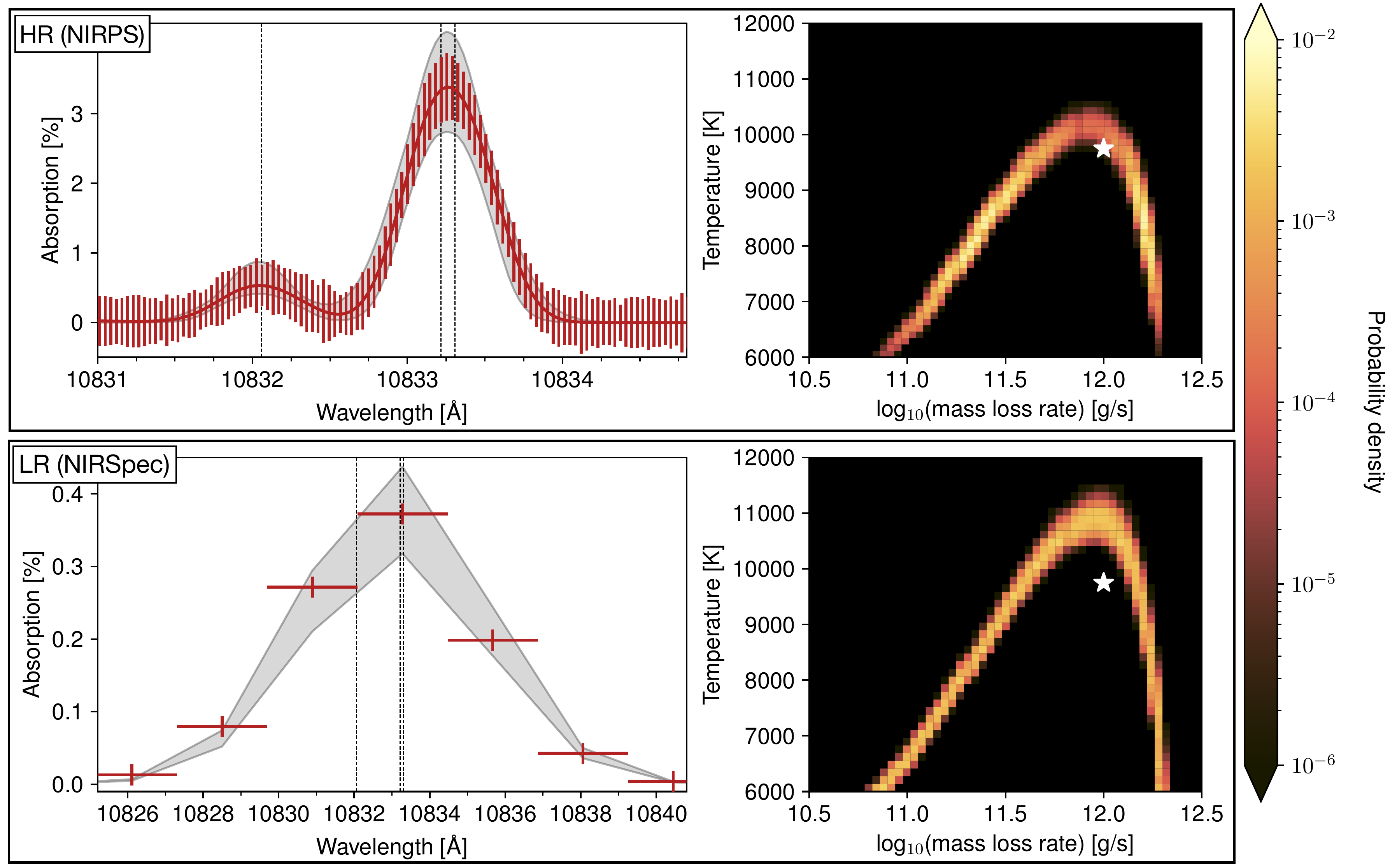}
    \centering
    \caption{Left panels: synthetic averaged absorption spectra for NIRPS (top) and NIRSpec (bottom) generated with \texttt{EvE}. The grayed area represents the 1~$\sigma$ envelop of all accepted models within the \texttt{p-winds} grid. The 3 vertical black dashed lines correspond to the helium triplet rest wavelengths. Right panels: distribution of atmospheric parameters found from the \texttt{p-winds} fit for NIRPS (top) and NIRSpec (bottom). The white star denotes the input parameters used to generate the averaged absorption spectra on the left panels.}
    \label{fig:convo}
\end{figure*}

Finally, we generate a set of 10\,000 mock observations from the \texttt{EvE} synthetic average absorption spectrum, and fit each of them with the \texttt{p-winds} grid, consistent with the common practice in the literature \citep[see e.g.][]{Kirk2022,allart2023,Vissapragada2024}. We retain all parameter sets within 1~$\sigma$ of the best fit, building a statistical distribution of the retrieved parameters, which can be interpreted as the probability of retrieving a specific set of parameters given the underlying planetary atmospheric structure. We show these results in Fig.~\ref{fig:convo} for both NIRPS and NIRSpec.

\smallbreak
\smallbreak
\noindent \textbf{High-resolution with NIRPS:}

Fig.~\ref{fig:convo} shows a degeneracy in the parameter space at HR. This degeneracy is mostly driven by the thermospheric model alone, as different combinations of mass-loss and temperature result in similar density profiles. 
Moreover, the statistical fluctuations in the observed flux within each spectral, bin due to photon noise, propagate the measurement uncertainties into variations in the depth and width of the absorption spectrum, resulting in a confidence band whose width depends on the S/N. However, this cannot account for the observed distribution here, since the input parameter set ends up not being the one most frequently recovered. We should also keep in mind that we considered a system with high S/N for the absorption signal. At lower S/N, the impact of this deviation on the derived parameters would be less significant and more easily mistaken with noise.

As shown in several previous HR studies \citep[see e.g.,][]{Casasayas2020,Dethier2023}, stellar disk inhomogeneities can strongly impact the HR absorption features. Notably, in the more realistic synthetic spectra we generated with \texttt{EvE}, the CLVs and RM induce POLDs along the transit chord. As a result, POLDs deform the absorption signature and, if left unmodeled, introduce a bias in the derived atmospheric parameters. Another important aspect is the 3D structure of the upper atmosphere implemented in \texttt{EvE}. Although we generated a purely 1D density profile with \texttt{p-winds}, it is projected into 3D. This allows us to account for the projected velocity field shifting the absorption feature. Not using a 3D description introduces a stronger bias for eccentric planets and large or complex atmospheres, whereas POLDs originate from the star and are stronger for fast rotators, cool host stars, and misaligned planets.

Despite these effects not being treated in \texttt{p-winds}, their impact on the atmospheric escape parameters is comparable to the HR instrumental uncertainties. As such, simplified radiative transfer approaches are sufficient as long as the relative precision on the absorption spectrum is not extremely high. However, for HR studies aiming for high precision, having high S/N on the absorption signal (i.e. strong absorption and/or high S/N on the stellar spectrum), or involving strong POLDs, we recommend using a more advanced radiative transfer model to properly isolate the planetary features from the POLDs \citep[see also][for a more general discussion]{Dethier2024}.

\smallbreak
\smallbreak
\noindent \textbf{Low-resolution with JWST:}

In the lower panels of Fig.~\ref{fig:convo}, we focus on LR with NIRSpec. Because of the broader instrumental kernel, absorption features become substantially shallower and wider, leading larger uncertainties in the retrieved parameters, as highlighted by the wider confidence band. This is naturally expected at LR, as the absorption feature is diluted by the instrumental kernel and spread across only a few spectral bins that are wider than the intrinsic width of the triplet. Consequently, the absorption width at LR is mainly determined by the instrumental convolution kernel, and information about the shape of the intrinsic absorption profile is lost. The resulting degeneracy between the width and the height of the profile translates into a degeneracy between mass-loss rate and temperature.

Furthermore, we note a stronger bias in the retrieved parameters compared to NIRPS. Yet, we expect the impact of neglecting POLDs and the 3D atmospheric structure to decrease at LR, as they induce spectrally narrow variations in the transmission spectrum that are smoothed out similarly to the absorption signal \citep{Carteret2024}. Moreover, for systems that are not strongly misaligned, \citet{Carteret2024} showed that the RM effect remains significantly below the observational uncertainties. Since the RM effect dominates the POLDs contribution over CLVs \citep{Yan2017,Casasayas2020,Dethier2023}, POLDs alone cannot explain the increased bias observed in the LR parameter space.

In this section, we reproduced the approach used in most studies in the literature, using a simplified radiative transfer treatment in which the stellar spectrum is ignored and absorption signatures are assumed to directly represent the transmission spectrum. As a result, the modeled absorption spectrum is directly convolved and resampled onto the observational spectral grid before being compared with the data. Yet, the transmission spectrum derived from the observations is computed as the ratio between the in-transit and out-of-transit flux spectra, according to Eq.~\ref{eq:abs_sp}. While convolution effects are not a major concern at HR, since the instrumental kernel is narrower than typical stellar lines, they can lead to non-negligible differences at lower resolutions. This explains why the input parameters are not correctly recovered. We further discuss the origin of this convolution bias in Appendix.~\ref{apdx:conv_bias}. Finally, we note that this conclusion is independent of the model used to compute the thermospheric density profile and originates only from the degradation of the absorption spectrum at LR.

\subsection{Predicting high-resolution absorption amplitude}
\label{sec:prediction}
To further illustrate the convolution bias at LR, we use our synthetic JWST LR observations, computed with Eq.~\ref{eq:abs_sp}, to predict the amplitude of the corresponding absorption spectrum at HR. In this way, we provide a first-order, purely data-driven estimate of the expected HR signal amplitude. This is particularly useful for assessing the detectability, from ground-based facilities, of signals detected with the JWST \citep[e.g. for sub-Neptunes,][]{ahrer2025} or to compare with previous HR observations. To this end, we approximate the HR absorption spectrum by a single Gaussian and compute mock NIRSpec observations using Eq.~\ref{eq:abs_sp} and the stellar spectrum, adopting the same uncertainties as in the previous section. We refer to this absorption spectrum as the true JWST absorption. We then fit the true JWST absorption using the HR Gaussian absorption profile after direct convolution and resampling (hereafter referred to as star-biased absorption), as is commonly done in the literature \citep[e.g.][]{Fournier-Tondreau2024,Piaulet-Ghorayeb2024,Radica2024,Allart2025Nat}. As shown in the previous section, this approach is subject to the convolution bias because the stellar spectrum is not included in the computation of the LR model absorption spectrum.

The true JWST mock absorption is generated assuming a peak absorption of 3\% and a full width at half maximum (FWHM) set to 0.7\,\AA, typical values found in atmospheric escape studies. We further consider two stellar spectra, shown in Fig.~\ref{fig:stellar_spectra}, that are representative of narrow and wide stellar lines. The first spectrum corresponds to WASP-69 and shows deep stellar absorption lines, whereas the second corresponds to WASP-121, an earlier-type star characterized by broader and shallower lines.

\begin{figure}
    \includegraphics[width=0.9\columnwidth]{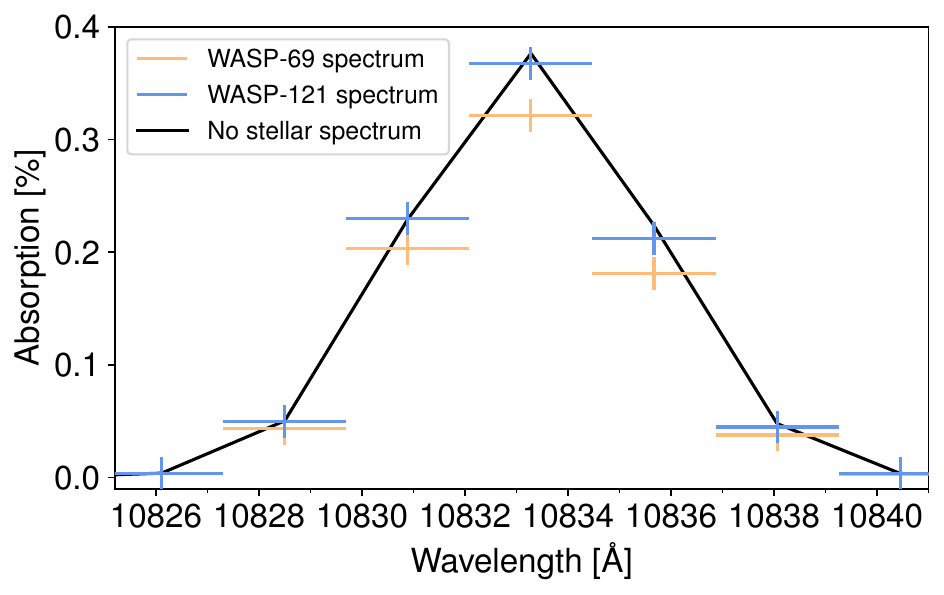}
    \centering
    \caption{Absorption spectra at NIRSpec resolution. The blue and orange spectra are computed correctly using Eq~\ref{eq:abs_sp}, from two different reference stars and a Gaussian high-resolution absorption with fixed FWHM of 0.7\,\AA and height of 3\%. The black spectrum is computed convolving directly the high-resolution Gaussian absorption, rather than using the ratio of the convolved in- and out-transit spectra (star-biased).}
    \label{fig:convo_gauss}
\end{figure}

We show in Fig.~\ref{fig:convo_gauss} the true JWST NIRSpec absorption spectra for the two stellar spectra.  
To further quantify this bias, we fixed the FWHM at its reference value in the fit to the two true NIRSpec spectra. We found the values $A_{69} = 2.52\pm 0.09 \%$ and $A_{121} = 2.93\pm 0.09 \%$ for the height of the Gaussian absorption spectrum, set originally to 3\%. We immediately see that the star-biased approach introduces a significant deviation ($\sim 4\,\sigma$ at the peak) compared to the true JWST absorption spectrum generated using WASP-69 stellar spectrum. Using the fitted values, this corresponds to a 16\% relative deviation from the expected value, or $\sim 1.5 \sigma$ for one NIRPS WASP-69 transit. We conclude that neglecting WASP-69 stellar spectrum significantly affects the retrieved height of the signal at HR. However, when neglecting the WASP-121 spectrum, the bias is much smaller and the predicted amplitude remains consistent with the original value.

We attribute this difference in bias magnitude to the shape of the stellar lines near the triplet (see also Appendix.~\ref{apdx:conv_bias}). In particular, WASP-69 stellar lines are deep, whereas those of WASP-121 are shallower. For narrow and deep stellar lines, the LR convolution averages the varying stellar flux over a broad range. Planetary absorption reduces even more the stellar flux near the core of the line but the resulting relative difference in flux is much lower for deep stellar lines. As a result, once convolved, the in-transit flux is proportionally less reduced for deep and narrow stellar lines, and thus absorption features smaller than the instrumental kernel are reduced. Consequently, directly convolving the absorption spectrum leads to a significant overestimation of the LR absorption in the case of deep stellar lines, compared to computing it from Eq.~\ref{eq:abs_sp}. This can lead to inconsistencies between HR and LR absorption measurements, which could be wrongly attributed to variability in the atmospheric signal \citep[see also discussion in][]{Dethier2024}.

We also note that for thermospheric layers close to the planet, where the gas is co-moving with the planetary bulk such that line broadening is dominated by thermal motions with typical temperature of a few $10^3$\,K, fixing the FWHM at $\sim0.7$\,\AA is a reasonable assumption. However, for outflows that extend further than a few planetary radii, the dynamical broadening (orbital motion of the gas) becomes dominant as the gravitational binding to the planet weakens \citep[see e.g.][]{Bourrier2018_Pancet,Zhang2023,Gully_Santiago2024}. Another mechanism that can increase the FWHM is the saturation of the deepest atmospheric layers, creating broader wings than expected \citep[see e.g. predicted spectra for various sub-Neptunes][]{Allan2025}. In such cases, the absorption spectrum can broaden significantly \citep{Huang2023,Czesla2024,Nail_2025}, and fixing the FWHM a priori can bias the derived HR amplitude (see also Appendix~\ref{apdx:FWHM_prediction}).

\subsection{Broadband molecular features}
\label{sec:broadband}

We previously discussed the convolution bias in the specific context of spectrally localized absorption features. The JWST, and LR observations more generally, are primarily used to probe broadband molecular features in exoplanet atmospheres. We therefore also quantify the bias introduced by neglecting the stellar spectrum when computing molecular absorption spectra over a broad wavelength range. For this, we simulated with \texttt{SCARLET} \citep{Benneke2012,Benneke2013} the transmission spectrum of WASP-69\,b at $\mathcal{R} = 250\,000$, assuming a constant temperature profile and solar abundances. Then, we computed the LR absorption spectrum as observed by NIRISS in three different ways: (REF) using the stellar spectrum and equation~\ref{eq:abs_sp}, (A) by directly resampling the \texttt{SCARLET} absorption spectrum onto NIRSpec wavelength table without convolution, and (B) by directly convolving and resampling the \texttt{SCARLET} absorption spectrum. 

\begin{figure}
    \includegraphics[width=0.9\columnwidth]{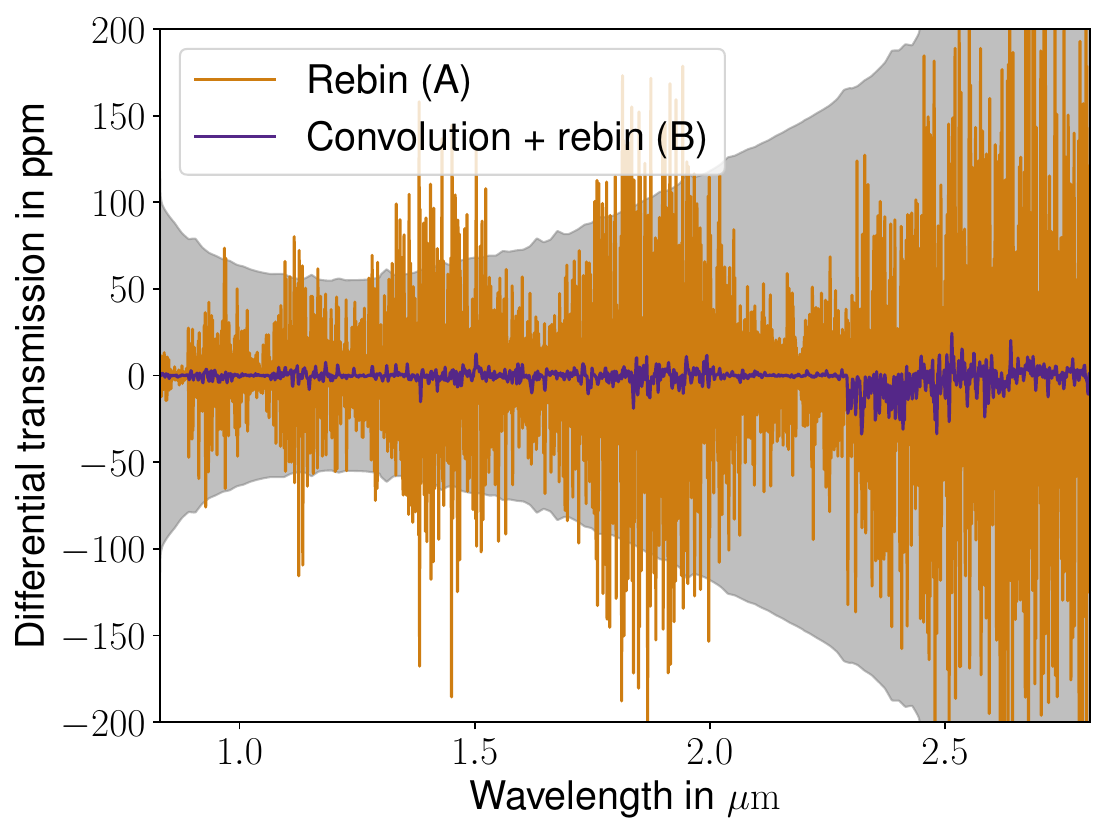}
    \centering
    \caption{Differential transmission spectra at NIRISS/SOSS resolution. We subtracted the absorption spectrum computed with equation~\ref{eq:abs_sp} to the direct resampling (brown) or convolution+resampling (purple) of WASP-69\,b simulated absorption spectrum. Error-bars for 1 NIRISS/SOSS transit vary between 50 and 250 ppm over the wavelength range (shaded area).}
    \label{fig:convo_molec}
\end{figure}

To estimate the significance of these deviations with respect to typical uncertainties in the absorption spectrum, we also computed the error-bar level associated to one transit of WASP-69\,b with NIRISS using \texttt{Pandexo}. Then, we derived two differential transmission spectra, defined as the differences between the incorrectly computed absorption spectra (A and B) and the correctly computed one (REF). We show in Fig.~\ref{fig:convo_molec} the two differential transmission spectra obtained for WASP-69\,b. 

We find that applying only spectral resampling (A) leads to large deviations relative to the correctly computed JWST absorption spectrum (REF). The amplitude of this deviation seems to increase with wavelength, which may be explained by the increasing NIRISS resolving power (see also Sect.~\ref{sec:prediction}). When we directly convolve and resample the absorption spectrum (B), we still observe some deviations but much smaller. Errors vary between 50 and 250 ppm depending on wavelength. While the amplitude of the bias exceeds uncertainties for the resampling alone (A), the resampling and convolution (B) bias remains much smaller. This contrasts with the previous section, where we observed deviations several times larger than the error-bars. We explain this by the spectral width of the molecular features (which are a blend of many individual lines), larger than both the convolution kernel and the stellar lines' width, resulting in a much lower impact from the stellar spectrum than for individual lines (see also Appendix.~\ref{apdx:conv_bias}). We note, however, that depending on the stellar type\footnote{See Fig.~\ref{fig:convo_molecW121} for the same computation with WASP-121 stellar spectrum.}, and amplitude of the absorption features, the bias level can vary and may become comparable to the error-bars. Based on our results, we strongly encourage LR retrieval frameworks to at least apply instrumental convolution, and ideally to use Eq.~\ref{eq:abs_sp} to correctly compute the model transmission spectrum while accounting for the stellar spectrum. Finally, as shown previously by \citet{Deming2017}, the spectral sampling of the stellar spectrum must be high enough to resolve properly all stellar lines in order to avoid any bias.

\section{High versus low-resolution escape detection and characterization}
\label{sec:high_vs_low}

In this section, we investigate and characterize the performance of HR and LR instruments in detecting and constraining atmospheric escape. Recently, the JWST has been used to detect atmospheric escape through the He\,\textsc{i} line, mainly with the NIRISS instrument, as a secondary product of transit observations \citep[e.g.][]{ahrer2025,Allart2025Nat,Fournier-Tondreau2025,Krishnamurthy2025}. JWST also covers the He\,\textsc{i} with the NIRSpec instrument, which has an intermediate resolution between that of NIRISS and ground-based spectrographs. With these different LR and HR instruments available to the community, it becomes essential to characterize the capability of each of them to detect He\,\textsc{i} in order to guide future observing programs. 

\subsection{Detection threshold}
\label{sec:detect_threshold}

\begin{figure*}[!h]
    \includegraphics[width=1.9\columnwidth]{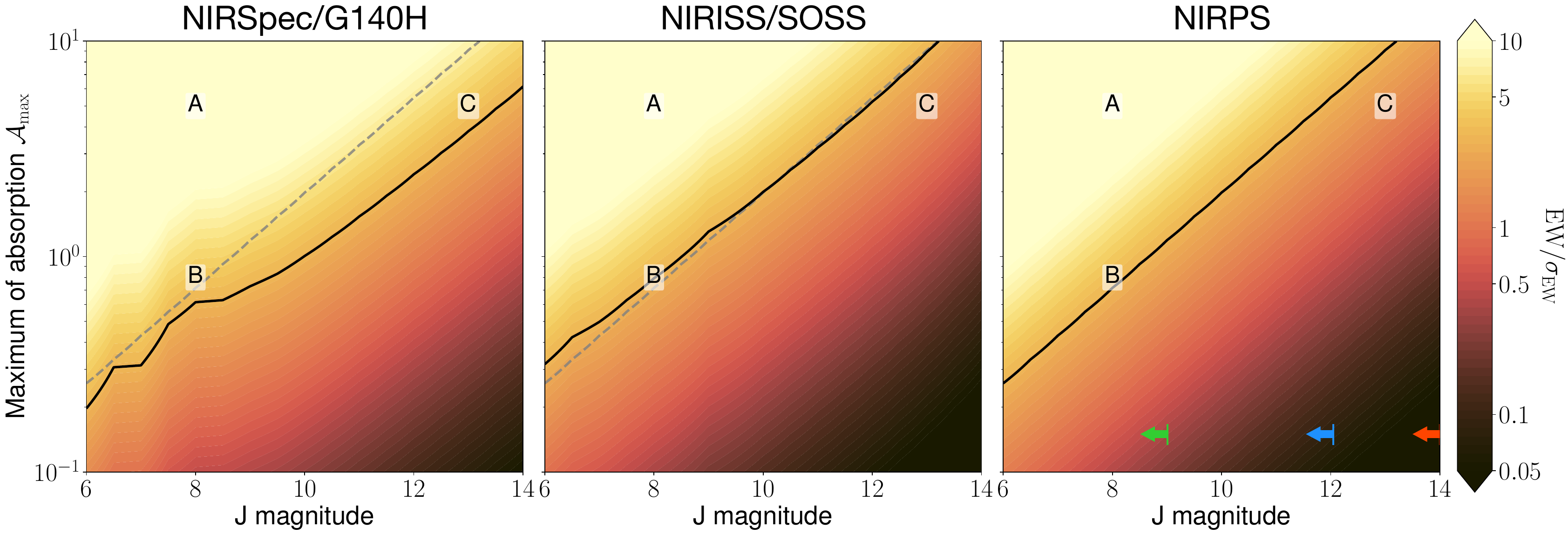}
    \centering
    \caption{Detection maps of He\,\textsc{i} absorption for NIRISS/SOSS, NIRSpec/G140H, and NIRPS. The black lines denotes the $3\sigma$ limit on $\mathrm{S/N}_\mathrm{EW}$, and we highlighted 3 specific cases with letters for which the observed absorption spectra are displayed in Fig.~\ref{fig:abs_features}. The dashed line reported on each JWST maps is the NIRPS $3\sigma$ threshold for comparison. On the NIRPS map we also show with the green, blue, and red arrows the J magnitude limit for $3\sigma$ characterization of the local planetary continuum for a typical sub-Neptune, Neptune, and Jupiter sized planet.}
    \label{fig:sensitivity_maps}
\end{figure*}

We generated a typical absorption signature with \texttt{p-winds}, using the WASP-69\,b parameters. We then scaled the absorption spectrum so that the peak excess absorption reaches a set of desired value. We computed the mid-transit flux spectrum associated with each of these model absorption spectra using the WASP-69 stellar spectrum, and used Eq.~\ref{eq:abs_sp} to properly compute the observed absorption spectrum. Finally, we estimated error-bars as described in Sect.~\ref{sec:method}, assuming a single transit with a baseline equal in duration to the transit. 

We use the excess absorption EW and associated error as a proxy for He\,\textsc{i} detections. Indeed, the EW fully integrates the absorption signature, which ultimately allows for a cross-comparison between instruments, and represents a good estimate of the mass-loss \citep{Ballabio2025}. However, because the absorption needs to be computed from Eq.~\ref{eq:abs_sp} and not by convolving directly the absorption spectrum (we discuss this aspect further in Sect.~\ref{sec:instrumental_convolution}) the EW is not comparable between the three instruments. As such, we adopt $\mathrm{S/N}_\mathrm{EW} = \mathrm{EW}/\sigma_\mathrm{EW}$ as a natural criterion to assess He\,\textsc{i} detections, setting a detection threshold at $\mathrm{S/N}_\mathrm{EW}\geq 3$. The EW integration ranges depend on the resolving power and are defined such that the entire absorption is covered. Furthermore, we considered that the continuum level, which originates from absorption by the planet opaque layers, is estimated locally around the He\,\textsc{i} lines. For HR we used a normalizing band of $\sim 10\,\mathrm{\AA}$ to avoid high frequency contamination from Earth's atmosphere and systematic noise, whereas we used $\sim 150\,\mathrm{\AA}$ for LR to correct for local molecular absorption. Uncertainties on the EW are propagated from the errors on the absorption spectrum and the locally evaluated continuum.

We estimated the $\mathrm{S/N}_\mathrm{EW}$ of each instrument for different J magnitudes to create detection maps, shown in Fig.~\ref{fig:sensitivity_maps}. These maps inform us of the level of absorption required to claim a detection as a function of the system brightness. Because the absorption signature is significantly smoothed by the large convolution kernel of NIRISS, the $\mathrm{S/N}_\mathrm{EW}$ does not scale linearly with the sensitivity. In practice, both instruments detect essentially similar levels of absorption in the underlying spectrum, with the difference remaining marginal despite the much higher precision of NIRISS compared to NIRPS. 
NIRSpec, on the other hand, appears to be a good trade-off between sensitivity and spectral resolution, improving detection levels especially for the faintest targets. We attribute the difference between NIRISS and NIRSpec to the presence of the stellar Si line near the He\,\textsc{i} ($\sim3\,\mathrm{\AA}$ away). Due to its lower resolving power, NIRISS ($\sigma_\mathcal{R}\sim7\,\mathrm{\AA}$ near the triplet) fully blends the stellar Si and He\,\textsc{i} lines in contrast to NIRSpec ($\sigma_\mathcal{R}\sim2\,\mathrm{\AA}$ near the triplet). When we compute the true JWST absorption spectrum with Eq.~\ref{eq:abs_sp}, the blending of He\,\textsc{i} and Si in the flux spectra reduces the strength of the He\,\textsc{i} absorption. 
The two JWST maps also depends on the stellar spectral type (see also Sect.~\ref{sec:instrumental_convolution}) which is further linked to the relative widths and depths of the Si and He\,\textsc{i} lines. We note also that these maps depend on the exact shape of the He\,\textsc{i} absorption profile, especially its width, which remains fixed at $\sim0.7\,\mathrm{\AA}$ here. An increase in the width of the He absorption lines, which scales with the EW, will ultimately increase the $\mathrm{S/N}_\mathrm{EW}$ and lower the detection threshold, since the uncertainty on the EW depends primarily on the J magnitude (see Fig.~\ref{fig:sensitivity_maps_subNept}). However, this is system dependent, and the conclusions drawn here, based on the relative levels between instruments, remain valid. 

Finally, we highlighted three specific regimes on the maps, corresponding respectively to high absorption and high S/N (A), low absorption and high S/N (B), and high absorption and low S/N (C). These three cases are illustrated in Fig.~\ref{fig:abs_features}, where we display the synthetic absorption spectra generated with \texttt{EvE} as observed by each instrument. The NIRPS absorption spectra are very noisy for faint targets, or low amplitude absorption, making the features hard to distinguish from the noise. However, the maps show that NIRSpec allows clear detections for faint targets that would be missed by NIRPS or NIRISS (case C). In Fig.~\ref{fig:sensitivity_maps}, we further show on the NIRPS map three colored arrows representing the magnitude threshold above which the continuum is determined at significance $<3\sigma$ for typical sub-Neptune, Neptune, and Jupiter planets. We note that both JWST instruments reach this threshold for J magnitudes greater than 14 for all planets. This highlights that, although JWST benefits from a considerably larger collecting area than the ESO 3.6m telescope, the high spectral resolution of NIRPS compensates for this difference in the case of narrow, localized features such as the He\,\textsc{i}. The two facilities therefore achieve comparable S/N on the helium integrated absorption, and thus yield similar detection thresholds.

\subsection{HR and LR thermosphere constraints}

\begin{figure*}
    \includegraphics[width=1.6\columnwidth]{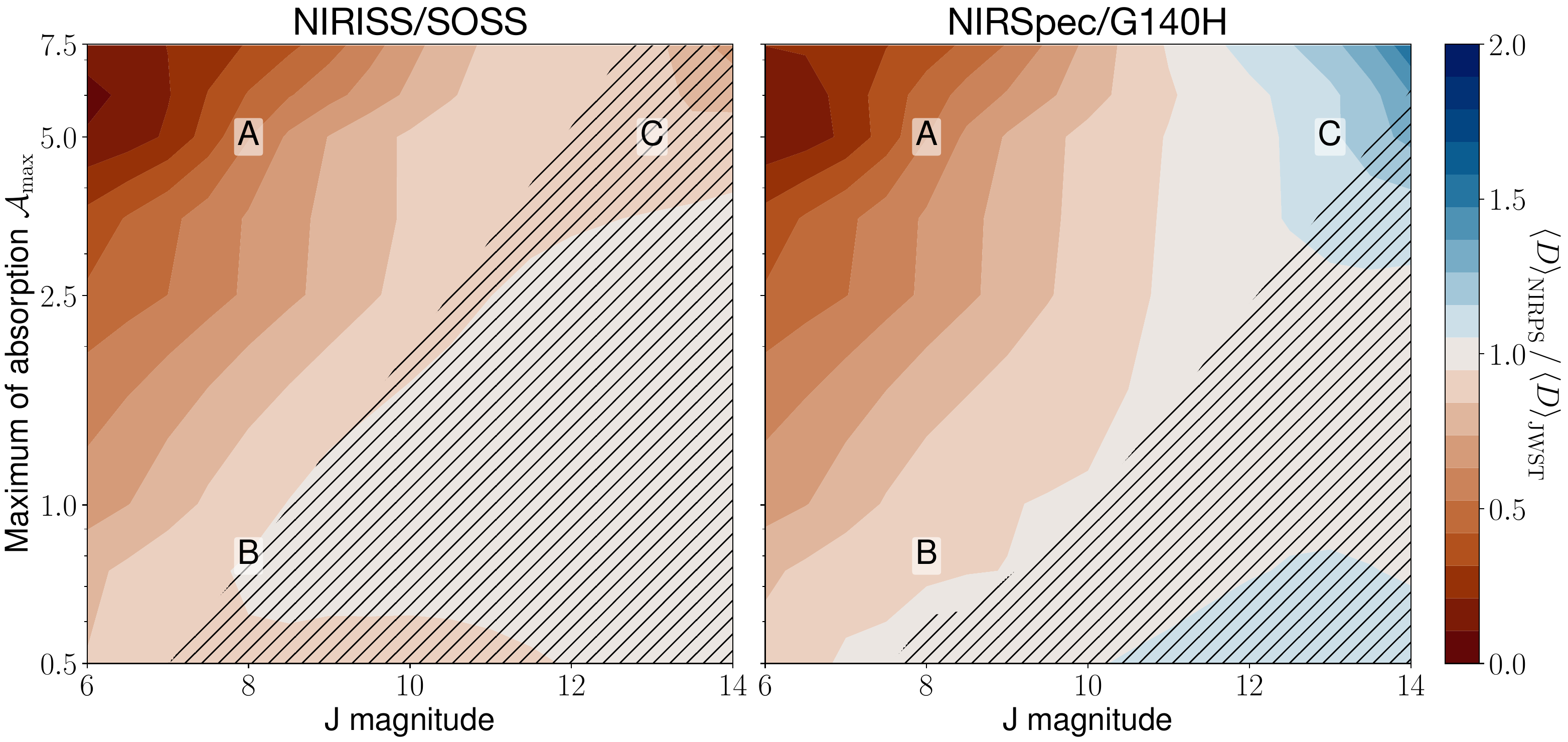}
    \centering
    \caption{Atmospheric escape fitting constraints maps for NIRISS/SOSS and NIRSpec/G140H with respect to NIRPS. We show in the maps the ratio of the $\langle D \rangle$ metric (Eq.~\ref{eq:d_metric}) between NIRPS and a JWST instrument (NIRISS left and NIRSpec right). Red colors indicate regions where NIRPS performs better than JWST in constraining the mass-loss - temperature parameter space, while blue ones indicate the opposite. The black hash-lines report the non-detection parameter space derived in Fig.~\ref{fig:sensitivity_maps}. Similarly to Fig.~\ref{fig:sensitivity_maps}, we also highlighted with letters the 3 specific cases for which the observed absorption spectra are displayed in Fig.~\ref{fig:abs_features}. }
    \label{fig:sensitivity_metric}
\end{figure*}

In the previous section, we focused on the capability of each instrument to detect atmospheric escape signatures. We showed that NIRPS and NIRISS spectra become less sensitive for faint targets, resulting in weak detections of atmospheric escape signatures. In this section, we aim to identify which instrument provides the most valuable information to characterize the upper expanding atmosphere. To this end, we generated a set of mock absorption spectra with \texttt{p-winds} using Eq.~\ref{eq:abs_sp}, varying temperature and mass-loss to obtain specific values of amplitude ($\mathcal{A}_\mathrm{max}$), while ensuring that the FWHM is preserved. Finally, we computed uncertainties for different J magnitudes as described in Sect.~\ref{sec:method}.

We then fitted each mock absorption spectrum using \texttt{p-winds} to derive constraints on temperature and mass-loss rate for each of these mock spectra. Our goal is to determine which instrument yields the strongest constraints on the atmospheric escape structure, as inferred from the absorption spectrum, for different absorption amplitudes and J magnitudes. The choice of \texttt{p-winds} arises naturally, as it offers a simplified but well-defined set of fitting parameters compared to 1D hydrodynamical codes \citep[e.g.][]{Caldiroli2021,Schulik2023,linssen2024} which are inherently less flexible. Nevertheless, we emphasize that our approach remains generic and should not depend on the choice of the model, since our analysis relies on a cross-comparison between instruments.

In order to evaluate the level of constraint provided by each instrument in the temperature - mass-loss parameter space, we computed the weighted average distance $\langle D \rangle$ to the input parameters for each pair of ($\mathcal{A}_\mathrm{max}$, J magnitude):
\begin{equation}
\label{eq:d_metric}
    \langle D \rangle_{\alpha,\beta} =  \frac{\sum_{i,j}\omega_{i,j}(\beta) \sqrt{(\frac{T_i-T_\alpha}{\Delta T})^2 + (\frac{\log{\dot{M}}_i-\log{\dot{M}}_\alpha}{\Delta \log{\dot{M}}})^2}}{\sum_{i,j}\omega_{i,j}(\beta)},
\end{equation}
where $T$ and $\log{\dot{M}}$ are sampled uniformly, $\alpha$ indexes a given $\mathcal{A}_\mathrm{max}$, $\beta$ indexes a given level of errors on the absorption spectrum, $\Delta T$ and $\Delta \log{\dot{M}}$ are the ranges within which each parameter is varied. Distances are weighted by the likelihood of their associated models:
\begin{equation}
    \omega_{i,j}(\beta) = \exp{(-\frac{1}{2}\chi(\beta)^2 )}.
\end{equation}
The metric $\langle D \rangle$ naturally captures the statistics of the parameter space, yielding smaller values for more constraining fits. We note that this approach is in principle degenerate between precision and accuracy, as a low precision but accurate or a precise but inaccurate fit could return the same $\langle D \rangle$. Furthermore, an intrinsic degeneracy of the parameter space arises because multiple combinations of temperature--mass-loss rate can produce nearly identical absorption spectra in \texttt{p-winds} (as discussed in Sect.~\ref{sec:convolution_thermosphere}). However, since we only add error-bars to the \texttt{p-winds} absorption spectra, the set of $T_\alpha,\log{\dot{M}}_\alpha$ always minimizes the $\chi^2$ value. Consequently, the metric probes the intrinsic degeneracy of the parameter space for a given absorption signature, rather than the accuracy of the best fit. This effect is system dependent and does not reflect the sensitivity of the instruments themselves. To mitigate this, we restrict our analysis to the ratio of the metric between two instruments, thereby filtering out the intrinsic parameter space degeneracy.

In Fig.~\ref{fig:sensitivity_metric}, we present the ratio between NIRPS and the two JWST instruments discussed throughout this paper. Red regions indicate where NIRPS provides stronger constraints than JWST, whereas blue regions indicate the opposite. It is worth noting that non-detections can still provide constraints on the temperature--mass-loss parameter space \citep{allart2023,masson2024}, and are therefore retained and reported in the figure. We find that NIRISS consistently provides weaker or equivalent constraints than NIRPS. This was expected, as convolving to LR smooths the absorption spectrum, leading to a degeneracy between width and amplitude. In our case, this translates to a mass-loss and temperature degeneracy which limits the precision of NIRISS fits across the parameter space. 
As shown in the previous section, NIRISS increased sensitivity (compared to NIRPS) does not compensate for the blurred, and thus reduced, absorption amplitude, nor does it mitigate the degeneracy of the parameter space. In contrast, we find that NIRSpec increased sensitivity and intermediate resolving power are sufficient to better constrain the parameter space than NIRPS for faint targets. Notably, NIRPS outperforms both JWST instruments for J<11, while NIRSpec consistently provides tighter constraints than NIRISS over all magnitudes.

Similarly to the previous section, we highlight three regimes of amplitude and S/N of absorption to further illustrate our conclusions. Case C detection is achievable only with NIRSpec, whereas cases A and B can be probed by all instruments. In case A, NIRPS provides the best constraints, while case B yields comparable performance across all instruments. We note, however, that a complete characterization of a system should also include the analysis of the light curve, as observations have shown that atmospheric escape can produce non-spherical thermospheres \citep{Allart2025,Krishnamurthy2025}. As discussed in Sects.~\ref{sec:extended_outflow}~\&~\ref{sec:baseline}, JWST instruments are much more efficient than ground-based facilities for probing extended and asymmetric outflows but do not allow for a detailed study of the outflow kinematics. Several He\,\textsc{i} detections have shown significant velocity shifts \citep{masson2024,Orell-Miquel2024}, which would not be resolved by JWST instruments. While it may not change the derived mass-loss rate by orders of magnitude, it does limit JWST's capability to investigate underlying physical processes. We discuss these aspects further in the next section.

\subsection{Extended outflow characterization}
\label{sec:extended_outflow}

In this section, we compare the performance of HR and LR for the characterization of extended outflows. To this end, we generated four different outflows with \texttt{EvE}, as observed with NIRPS and JWST. We show in Fig.~\ref{fig:shape} the four thermospheric shapes considered here, together with their respective exospheric contribution (see also Fig.~\ref{fig:maps} for the simulated absorption time series). The first three models are based on \citet{MacLeod2022}, who simulated hydrodynamical outflows interacting with the stellar wind, with each case representing an approximation of the planetary outflow confinement by stellar winds of different strengths: weak (W), medium (M), and strong (S). The associated thermospheres are populated using \texttt{p-winds} profiles with the same set of parameters. In all cases, the exosphere launch distribution depends on the strength of the stellar wind, which affects the global direction of the outflow. The last model approximates a massive Roche lobe overflow that forms dense streams \citep[STR,][]{macleod2024streams}. The four simplified models are motivated by simulations \citep{Wang2021,macleod2024streams} and recent observations \citep{Zhang2023,Gully_Santiago2024,Allart2025,Allart2025Nat,Krishnamurthy2025} of hydrodynamic He\,\textsc{i} outflows extending far from the planet. We note that these typical outflow shapes may also be influenced by the sound speed, the temperature of the outflow, or tidal interactions with the host star \citep{Nail2024,Nail_2025,macleod2024streams} but our four models broadly cover the range of structures that can be observed.

\subsubsection{Averaged spectra}

\begin{figure*}
    \includegraphics[width=1.8\columnwidth]{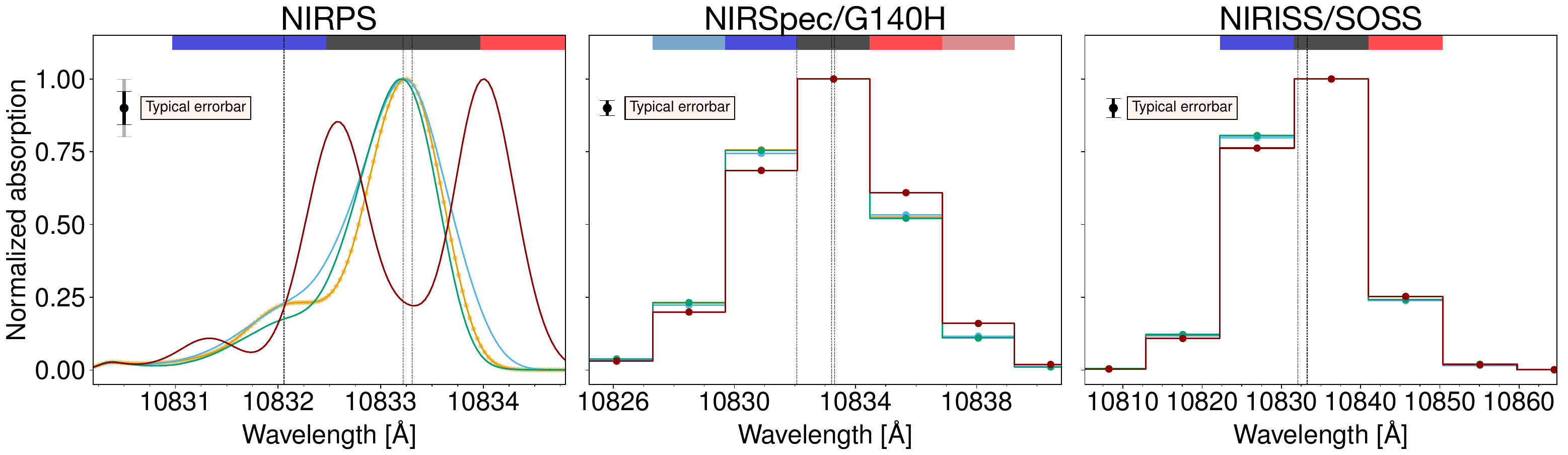}
    \centering
    \caption{Simulated average absorption spectra in the planetary rest-frame for the four different outflows considered (the color code used is the same as in Fig.~\ref{fig:shape}), as observed with NIRPS ($\mathcal{R} \sim 75\,000$, left),  JWST/NIRSpec ($\mathcal{R} \sim 2\,050$, middle) and JWST/NIRISS ($\mathcal{R} \sim 650$, right). The dots represent the centers of the instrumental spectral bins. In the NIRPS panel, black error-bars are shown for 3 NIRPS transits whereas the faded ones for a single one. The three black vertical lines show the positions of the helium triplet lines at rest. We highlighted on top of each panel the blue, red, and black regions which correspond to the spectral ranges used for the light curves computation (Fig.~\ref{fig:abs_lc_panels}).}
    \label{fig:abs_sp_panels}
\end{figure*}

The most commonly used metric, especially in high-resolution atmospheric escape studies, is the absorption spectrum in the planetary rest frame averaged over the transit window. In Fig.~\ref{fig:abs_sp_panels}, we show the averaged spectra associated with the studied instruments and four outflow scenarios. We normalized each absorption spectrum by its maximum amplitude to focus on the relative spectral features rather than their absolute levels that depend on the exact He\,\textsc{i} density profile. There are clear differences between outflows when observed at HR that reflect both the shape of the evaporating atmosphere and its specific dynamics. In the STR scenario, He\,\textsc{i} atoms are no longer gravitationally bound to the planet and thus do not follow its orbital motion, leading to strong blue (resp. red) shifts for the trailing (resp. leading) tail that are associated with two distinct peaks in the average absorption spectrum. We note also that, for a highly extended thermosphere (M case), the wings of the helium triplet are substantially broader at HR because the thermosphere is so extended that different regions have substantially different radial velocities (i.e. in projection along the LOS), enough to separate their absorption signatures. On the other hand, information about atmospheric dynamics at LR in JWST absorption spectra is limited due to spectral dilution and large pixel width. Thermal broadening of the thermospheric absorption signature is insufficient to distinguish between the three stellar-winds scenarios (W, M and S), while the STR scenario exhibits slight differences in the relative amplitudes of the spectral bins surrounding the triplet. 

Overall, the HR average absorption spectrum contains more information about the upper atmospheric dynamics than its LR counterpart, as expected. While LR spectra can only provide order-of-magnitude estimates of the LOS velocity, HR observations are crucial to properly study the dynamics of the upper atmosphere. However, this information is limited by the fact that the average spectrum is typically built from observations taken during the optical transit, between the second and third contact, which may not be representative of outflows extending far from the planet with asymmetric geometries and velocity fields. As a result, combined observations at LR from space (to probe the spatial extension of the outflow) and at HR from the ground (to probe its dynamics) are required to properly characterize a He\,\textsc{i} absorption time series.

\subsubsection{Light curves}

\begin{figure*}
    \includegraphics[width=1.75\columnwidth]{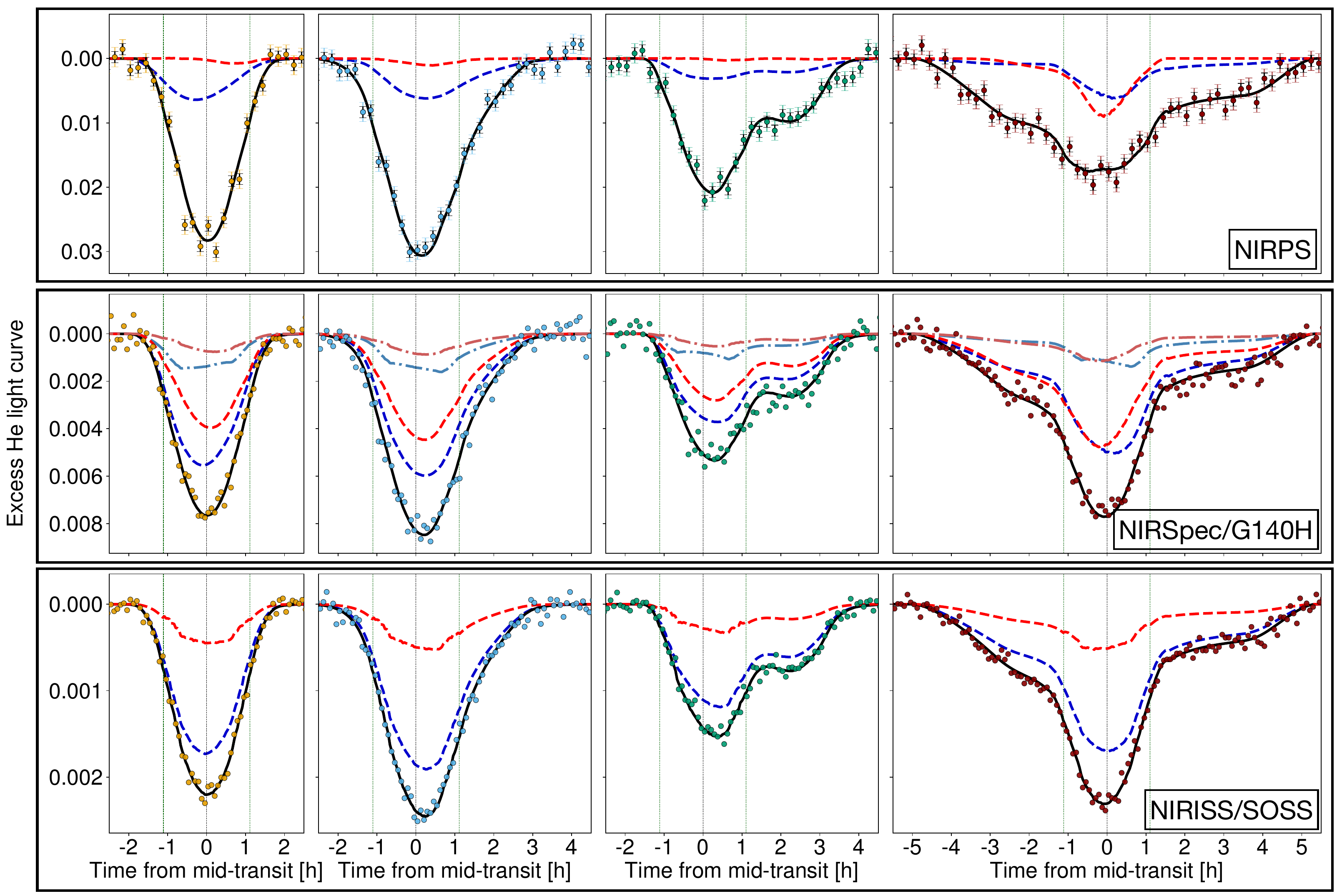}
    \centering
    \caption{Simulated excess absorption light curves in the stellar rest-frame for the four different outflows considered, as observed with NIRPS ($\mathcal{R} \sim 75\,000$, top),  JWST/NIRSpec ($\mathcal{R} \sim 2\,050$, middle) and JWST/NIRISS ($\mathcal{R} \sim 650$, bottom). The solid lines indicate light curves built from the black spectral range in Fig.~\ref{fig:abs_sp_panels}, where as blue/red dashed lines are built from the blue/red ranges of Table~\ref{tab:error}. The dots represent one random instance of observation for each cases. In the NIRPS panel, black error-bars are shown for 3 NIRPS transits whereas the faded ones for a single one. The three vertical lines show the first contact, mid-transit, and last contact times. The color code used for the light curves is the same as in Fig.~\ref{fig:shape}.}
    \label{fig:abs_lc_panels}
\end{figure*}

We present in Fig.~\ref{fig:abs_lc_panels} the simulated excess light curves of He\,\textsc{i} for the four different scenarios considered in this section. Each simulated light curve is shown alongside one realization of the observations, whose scatter is computed from the S/N as described in Sect.~\ref{sec:method}. The excess absorption observed in the NIRPS light curves is much stronger than in JWST observations due to spectral dilution caused by the instrumental convolution and resampling. 
However, ground-based observations face significant limitations such as the loss of the absolute flux level, limited observing windows due to night duration, and low temporal cadence. These constraints become particularly important for extended absorption signatures, e.g. in the STR scenario the helium transit can last longer than the night duration. However, stitching together data taken on different nights, sometimes separated by several days \citep{Zhang2023, Gully_Santiago2024}, requires coordination and is sensitive to variations in stellar activity level between nights. The most critical issue is the loss of the true flux baseline. In the context of atmospheric escape, absorption may begin long before the optical transit and extend well after. Combined with the low temporal cadence required to achieve sufficient S/N from the ground, potential telluric contamination or instrumental instabilities, this makes it particularly challenging to identify the start and end of the helium absorption signal and to define an appropriate observing strategy that covers pre- and post-transit baselines (see also Sect.~\ref{sec:baseline}). 
This limited cadence also reduces the ability to detect short-timescale variability in the escaping atmosphere.
In contrast, space-based instruments are not limited by the night duration and provide a much better baseline coverage. Furthermore, JWST light curves achieve a relative scatter comparable to that of NIRPS while maintaining a finer temporal sampling, which is crucial to reduce the uncertainty in the outflow extent, a key parameter to constrain the mass-loss. JWST has already demonstrated its value in revisiting known systems with new detections of pre- and post-transit absorption features \citep[e.g.][]{Allart2025Nat,Krishnamurthy2025}. Since LR observations allow for a better temporal sampling and time coverage than HR ones, they are essential for complementing HR observations to constrain the duration of absorption, which in turn allows for a better characterization of the outflow extent.

\section{Choice of baseline for atmospheric escape observations}
\label{sec:baseline}

HR transmission spectroscopy relies mostly on the comparison between spectra collected during the transit phases and the unabsorbed stellar spectrum, directly measured before and/or after the transit event. As pointed out in the previous section, in practice the identification of the baseline can be difficult with ground-based facilities, especially in the context of atmospheric escape where absorption may last significantly beyond the expected optical transit \citep[e.g.][]{Lavie2017}. 

Let us consider the true stellar spectrum given by $F_\star = \sum_{i=0}^n F_i$ on a discretized stellar grid, where ${F_i}$ stands for the local stellar spectra (see also Sect.~\ref{sec:instrumental_convolution}). When the stellar disk is occulted, the local stellar flux received by the observer is given by $F_{\mathrm{obs,i}}=F_i e^{-\tau_i}$, with $\tau_i$ the optical depth proportional to the cross-section and column density of the occulting species. Note that $\tau_i \rightarrow \infty$ for optically thick regions such as the rocky core or deep atmospheric layers. In the most generic way, the observed spectrum at a time $t_{\mathrm{obs}}$ is 
\begin{equation}
    F_\mathrm{obs}(t_{\mathrm{obs}}) = \sum_{i=0}^n F_i e^{-\tau_i}.
\end{equation}
Now consider the stellar spectrum computed from observations, referred to as the master spectrum. If the baseline is improperly identified, then the master may be biased by absorption from the extended atmosphere. In a general case, and assuming constant exposure times, the master spectrum is given by 
\begin{equation}
    F_\mathrm{master}=\frac{1}{N_\mathrm{bexp}}\sum_{\mathrm{bexp}} F_\mathrm{obs}(t_\mathrm{bexp}) ,
\end{equation}
where the subscript $\mathrm{bexp}$ refers to the baseline exposures. In the case where the baseline is correct this equation reduces to $F_\star$. However, if absorption occurs during the baseline, then the master spectrum becomes contaminated, which creates a bias in the entire absorption time series. The measured absorption at time $t_{\mathrm{obs}}$ is indeed given by 
\begin{align}
\label{eq:bias_mast}
\mathcal{A}(t_{\mathrm{obs}}) &= 1-\frac{F_\mathrm{obs}(t_{\mathrm{obs}})}{F_\mathrm{master}} \notag \\
&= \frac{\sum_{t_\mathrm{bexp}} (F_\mathrm{obs}(t_\mathrm{bexp}) / N_{\mathrm{bexp}}) - F_\mathrm{obs}(t_{\mathrm{obs}})}{\sum_{t_\mathrm{bexp}} (F_\mathrm{obs}(t_\mathrm{bexp}) / N_{\mathrm{bexp}})} \notag \\
&= \frac{\sum_{t_\mathrm{bexp}} (\sum_{i} F_i e^{-\tau_i(t_\mathrm{bexp})} / N_{\mathrm{bexp}}) - \sum_{i} F_i e^{-\tau_i( t_{\mathrm{obs}} )}}{\sum_{t_\mathrm{bexp}} (\sum_{i} F_i e^{-\tau_i(t_\mathrm{bexp})} / N_{\mathrm{bexp}})} \notag \\
&= \frac{\sum_{i} F_i  (\sum_{t_\mathrm{bexp}} e^{-\tau_i(t_\mathrm{bexp})} / N_{\mathrm{bexp}} - e^{-\tau_i(t_{\mathrm{obs}})})}{\sum_{i} F_i \sum_{t_\mathrm{bexp}} (e^{-\tau_i(t_\mathrm{bexp})} / N_{\mathrm{bexp}})} \notag \\
&= \frac{\sum_{i} F_i \big(\delta_i - e^{-\tau_i(t_{\mathrm{obs}} )}\big)}{\sum_{i} F_i \delta_i}
\end{align}
where all time sums are performed over the baseline and we define the quantity $\delta_i = \sum_{t_\mathrm{bexp}} e^{-\tau_i(t_\mathrm{bexp})} / N_{\mathrm{bexp}} $. If the baseline is well identified, then $\delta_i$ reduces to 1. However, if the baseline contains spectra where the outflow already absorbs the stellar flux, $\delta_i<1$ which propagates through the entire absorption time series, biasing it. Because $\tau_i$ is wavelength dependent, the bias also changes with wavelength. 

\begin{figure*}
    \includegraphics[width=1.82\columnwidth]{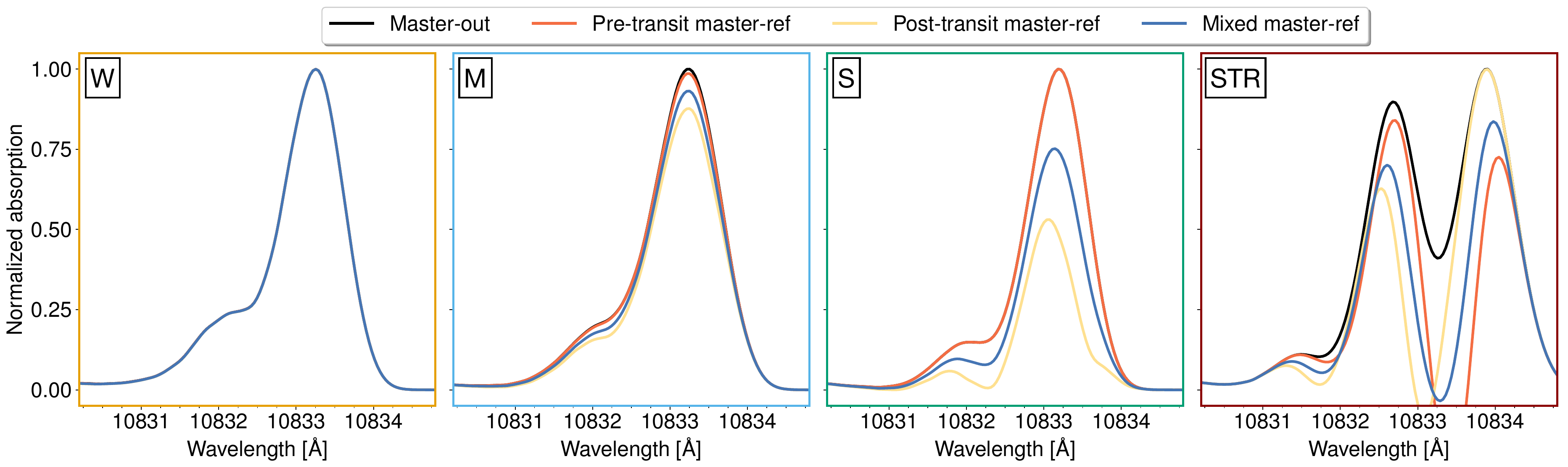}
    \centering
    \caption{Synthetic high-resolution average absorption spectra computed using various stellar reference spectra as observed with NIRPS ($\mathcal{R} \sim 75\,000$). All spectra are normalized by the peak of the absorption spectrum computed using the master-out (black) to help for comparison. }
    \label{fig:abs_baseline}
\end{figure*}

To illustrate the strength of this effect, we computed a biased master spectrum in our four simulation setups presented in Sect.~\ref{sec:extended_outflow}, by integrating the flux spectra over intervals from -3 to -2 hours and/or 2 to 3 hours, i.e. when the extended atmospheres is already transiting. In Fig.~\ref{fig:abs_baseline}, we show the averaged absorption spectra generated with the biased master spectra. In the fully spherical case, absorption occurs only within $\pm2$ hours so we do not see any difference with the true absorption spectrum. However, by looking at more extended upper atmospheres, we see that misidentifying the baseline underestimates the averaged absorption spectrum, as pointed out by equation \ref{eq:bias_mast}. Moreover, the bias varies with the line of sight velocity of the upper atmosphere. As such, direct comparisons between models and observations become more challenging, especially if we consider only the averaged absorption spectrum as a proxy.

In practice, all the cases illustrated in Sect.~\ref{sec:extended_outflow} can be distinguished using the light curves. However, the exact time at which absorption begins or ends is less straightforward to identify. Moreover, ground-based observations are limited by the night duration and the temporal sampling. These aspects make subtle variations or plateaus in the absorption harder to detect and easy to confuse with instrumental systematics or noise at HR. As an example, ground-based observations of the stream case would completely miss the stream structure if performed between -3\,h and 3\,h, a typical observation strategy for a transit duration of 2 hours. Thus, ground-based characterization of extended atmospheric escape requires the longest possible baseline coverage, especially for systems where streams or tails can be expected \citep[see predictions from][]{macleod2024streams}. In this regard, contemporaneous observations with JWST, and more generally with space-based facilities, are essential to properly isolate the planetary signature and extract mass-loss rates.

\section{Conclusions and Recommendations}
\label{sec:conclusion}

In this paper, we reviewed the different approaches to observing atmospheric escape through He\,\textsc{i}. In particular, we simulated different outflow configurations and generated mock observations for NIRPS, NIRSpec, and NIRISS. We discussed the strengths and weaknesses of ground- and space-based facilities in detecting and characterizing atmospheric escape. Finally, we illustrated how the choice of baseline can bias the absorption time series derived from the data.

Our study shows the importance of properly accounting for the stellar contribution and instrumental response when comparing absorption models and observations. At HR, the stellar CLVs and RM can mimic absorption or emission features when not taken into account, which prevents us from isolating the planetary signal \citep{Dethier2023}. Most models use the disk-integrated spectrum as a proxy for the local stellar spectrum, neglecting the POLDs. We find that this slightly impacts the retrieved parameters of the upper atmosphere for our representative system WASP-69\,b, but expect the bias from POLDs to increase for fast rotators and cooler host stars. At LR, \citet{Carteret2024} showed that POLDs do not bias the absorption in the helium band. On the other hand, we find that convolving the modeled absorption spectrum directly introduces a strong deviation when compared with LR measurements. The modeled absorption at LR must be computed using the stellar spectrum, as indicated in Eq.~\ref{eq:abs_sp}. Based on our results, we recommend always applying the instrumental convolution to the modeled flux spectra time series, rather than to the modeled absorption, when comparing models with observations, as implemented in \texttt{EvE}. This conclusion also applies to retrieval frameworks for broadband features.

Our comparison between NIRPS, NIRSpec, and NIRISS highlights that HR instruments (NIRPS) generally outperform LR instruments (JWST) for characterizing atmospheric escape through He\,\textsc{i}. NIRPS and NIRISS have comparable detection thresholds, whereas NIRSpec provides slightly improved detectability, especially for faint targets. Furthermore, contrary to \citet{DosSantos2023} we do not find that JWST instruments provide tighter (compared to ground-based spectrographs) constraints on atmospheric escape parameters over a broad range of absorption amplitudes. We note however, that NIRSpec seems to perform better than NIRPS for faint targets. By analyzing different outflow configurations, we would like to emphasize that HR and LR (space-based) measurements are fully complementary for studying atmospheric escape, despite probing the same layers. On the one hand, space-based LR instruments do not suffer from telluric or night duration restriction, and allow for a better temporal sampling. This is particularly important to constrain more precisely the duration of absorption, allowing for a better characterization of the outflow extent. A misidentification of the extent can bias the stellar spectrum used as reference both at LR and HR, which in turn affects the full absorption time series. On the other hand, HR observations are necessary to study the dynamics of the upper atmosphere, as LR can only provide an estimate of the LOS velocity. We recommend that future observing surveys maximize simultaneous observations with space- and ground-based facilities. One aspect we that did not discuss here is the temporal variability of the absorption signal which could prevent a proper comparison between HR and LR observations when taken too long apart.

\begin{acknowledgements}
This work has been carried out within the framework of the NCCR PlanetS supported by the Swiss National Science Foundation under grants 51NF40$\_$182901 and 51NF40$\_$205606. This project has received funding from the European Research Council (ERC) under the European Union's Horizon 2020 research and innovation programme (project {\sc Spice Dune}, grant agreement No 947634). 

\end{acknowledgements}

\bibliographystyle{aa}
\bibliography{biblio}

\begin{appendix}

\section{Convolution bias}

\subsection{Simplified example}
\label{apdx:conv_bias}
In this appendix, we introduce a simple mathematical framework to illustrate how the order in which convolution is applied and the flux ratio is computed can bias the absorption spectrum. Let us consider planetary absorption lines ($\mathcal{A}_{p,n}$), stellar absorption lines ($\mathcal{A}_{\star,n}$), and fluxes in- and out-of-transit ($f_\mathrm{in/out,n}$) at the intrinsic (i.e. infinitely high) resolution. To simplify the problem we assume the stellar surface to be homogeneous (neglecting POLDs), such that the local stellar intensity $I_i$ is constant across the surface element $S_i$: 
\begin{align}
    f_\mathrm{out,n} &= \sum_i I_i S_i \sim I S_\star = F_\mathrm{cont}(1-\mathcal{A}_{\star,n}), \\
    f_\mathrm{in,n} &= \sum_i I_i( S_i - S_{\mathrm{opaque}} - S_\mathrm{p}\mathcal{A}_{p,n,i}) \notag \\
    &\sim I S_\star \big( 1 - S_{\mathrm{opaque}}/S_\star - (S_\mathrm{p}/S_\star)\mathcal{A}_{p,n}\big) \notag \\
    & = F_\mathrm{cont}(1-\mathcal{A}_{\star,n})\big( 1 - S_{\mathrm{opaque}}/S_\star - (S_\mathrm{p}/S_\star)\mathcal{A}_{p,n}\big),
\end{align}
where $F_\mathrm{cont}$ is the stellar continuum. Note that $S_\mathrm{p}/S_\star$ is simply a geometrical factor and, since we arbitrarily fix the absorption amplitude, it can be absorbed into $\mathcal{A}_{p,n}$ for simplicity. We also set $S_{\mathrm{opaque}}=0$, assuming absorption is dominated by the extended atmosphere, and neglect spectral resampling. Using $G_t$ for the instrumental Gaussian kernel, absorption spectrum is then estimated as 
\begin{align}    
    \mathcal{A} &= \frac{f_\mathrm{out} - f_\mathrm{in}}{f_\mathrm{out}}, \notag \\
    &= \frac{f_\mathrm{out,n}\ast G_t - f_\mathrm{in,n}\ast G_t}{f_\mathrm{out,n}\ast G_t}, \notag \\
    & = \frac{(1-\mathcal{A}_{\star,n})\ast G_t - \big((1-\mathcal{A}_{\star,n})(1-\mathcal{A}_{p,n})\big)\ast G_t}{(1-\mathcal{A}_{\star,n})\ast G_t}, \\
    & = (\mathcal{A}_{p,n}\ast G_t) \frac{1-\frac{(\mathcal{A}_{\star,n} \mathcal{A}_{p,n})\ast G_t}{\mathcal{A}_{p,n}\ast G_t}}{(1-\mathcal{A}_{\star,n}) \ast G_t}, \notag \\ 
    & = (\mathcal{A}_{p,n}\ast G_t) \, \kappa. \notag
\end{align}

The advantage of this framework is that it explicitly relates the correctly computed observed absorption spectrum to the direct convolution of the native absorption spectrum. In this way, we can express the deviation introduced by the convolution simply by evaluating the $\kappa$ function, which depends on the planetary and stellar absorption profile and the instrumental kernel. Finally, we assume the planetary absorption lines, stellar absorption lines, and instrumental kernel to be Gaussian. Formally, we have 
\begin{equation}
    \mathcal{A}_{i}(\lambda) = A_i \exp{\big( -\frac{1}{2\sigma_i^2}(\lambda-\mu_i)^2\big),}
\end{equation}
where the subscript $i$ stands for planetary absorption ($p$), stellar ($\star$), or instrumental ($t$) kernels. Note that by definition $\mu_t = 0$, which is implicit for the rest of the section.

\begin{figure}
    \includegraphics[width=\columnwidth]{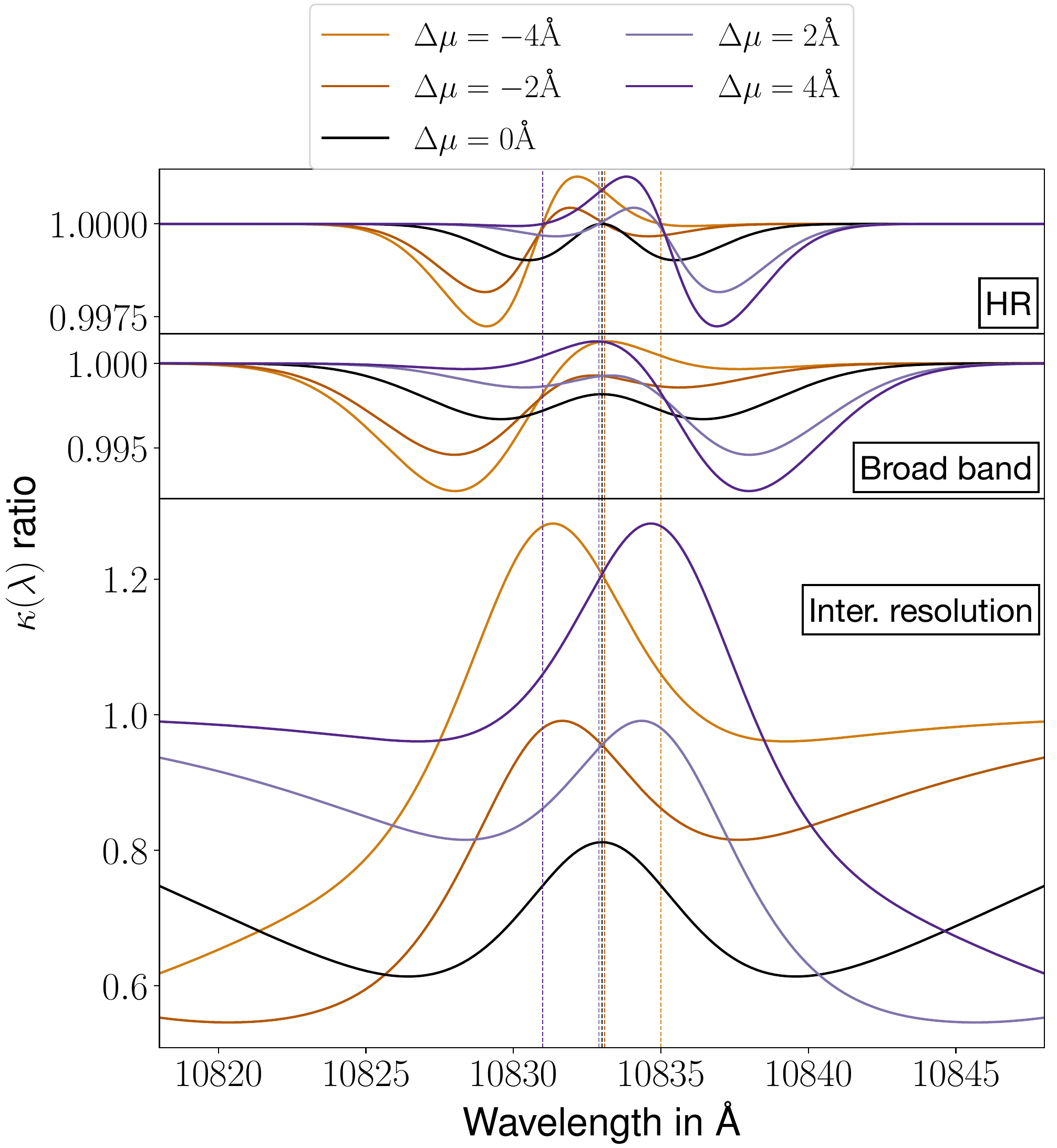}
    \centering
    \caption{Amplitude of $\kappa$ computed from Eq.~\ref{eq:kappa_gauss} for the three regimes discussed. The dashed lines correspond to the position of $\mu_p$ and we define $\Delta\mu = \mu_\star - \mu_p$. In all cases we fixed $A_p = 0.02$, $A_\star=0.5$, and $\sigma_\star=2 \mathrm{\AA}$. Top panel: High-resolution regime where $\sigma_p=1 \mathrm{\AA}$ and $\sigma_t=0.046 \mathrm{\AA}$ which corresponds to $\mathcal{R} = 100\,000$. Middle panel: Broadband absorption regime where $\sigma_p=20 \mathrm{\AA}$ and $\sigma_t=2.3 \mathrm{\AA}$ which corresponds to $\mathcal{R} = 2\,000$. Bottom panel: Intermediate-resolution regime where $\sigma_p=1 \mathrm{\AA}$ and $\sigma_t=2.3 \mathrm{\AA}$ which corresponds to $\mathcal{R} = 2\,000$. }
    \label{fig:gauss_toy_model}
\end{figure}

Under the previous simplifications, and using properties of Gaussian and convolution, we get 
\begin{equation}
\label{eq:kappa_gauss}
    \kappa(\lambda) = \frac{1-\frac{A_r}{A_p}\frac{\sigma_r}{\sigma_p}\sqrt{\frac{\sigma_p^2+\sigma_t^2}{\sigma_r^2+\sigma_t^2}} \exp{\bigg[ -\frac{1}{2} \big( \frac{(\lambda-\mu_r)^2}{\sigma_r^2+\sigma_t^2} - \frac{(\lambda-\mu_p)^2}{\sigma_p^2+\sigma_t^2} \big) \bigg]}}{1-A_\star\frac{\sigma_\star^2}{\sqrt{\sigma_\star^2+\sigma_t^2}} \exp{\bigg[ -\frac{1}{2} \big( \frac{(\lambda-\mu_\star)^2}{\sigma_\star^2+\sigma_t^2} \big) \bigg]}},
\end{equation}
where we introduced $\sigma_r^2 = \sigma_\star^2\sigma_p^2/(\sigma_\star^2+\sigma_p^2)$, $\mu_r = (\mu_\star\sigma_p^2+\mu_p\sigma_\star^2)/(\sigma_\star^2+\sigma_p^2)$, and $A_r = A_\star A_p\exp\big( -0.5(\mu_\star-\mu_p)^2/(\sigma_\star^2+\sigma_p^2) \big)$ that do not depend on instrumental kernel. This formula of $\kappa$ depends on the position of the lines, their amplitudes, and their width. We can now approximate and compare $\kappa$ in different regimes, shown in Fig.~\ref{fig:gauss_toy_model} and discussed below. 
\bigbreak
\noindent \textbf{High-resolution:} At HR, the width of the instrumental kernel is much smaller than the native absorption and stellar line. As such we have $\sigma_t \ll \sigma_p,\sigma_\star$ which leads to $\kappa(\lambda) = 1$. It follows that the stellar spectrum can be neglected in the computation of the absorption if POLDs are negligible. If that is not the case, as is generally true, the stellar spectrum is still required to accurately describe the local planetary absorption on the stellar disk. 
\smallbreak
\noindent \textbf{Broadband absorption:} If we consider molecular features, similarly to Sect.~\ref{sec:broadband}, the forest of narrow and shallow lines can roughly be approximated by an absorption feature much broader than both the stellar lines or instrumental kernel. In this regime, $\sigma_p \gg \sigma_t,\sigma_\star$ and $\kappa(\lambda)$ again converges to 1. This explains why the amplitude of the differential transmission spectrum in Sect.~\ref{sec:broadband} was smaller for molecular features than for localized absorption features (Sect.~\ref{sec:prediction}). 
\smallbreak
\noindent \textbf{Intermediate-resolution:} In the case of He\,\textsc{i} absorption at NIRSpec resolution we have roughly $\sigma_p \sim \sigma_t \sim \sigma_\star$. This time, $\kappa(\lambda)$ does not simplify to 1, and depends on the relative position of the stellar and absorption lines. If we assume for simplicity $\mu_p = \mu_\star$, corresponding to mid-transit, and evaluate $\kappa$ at $\lambda=\mu_p$ then we obtain 
\begin{equation}
    \kappa(\lambda) = \frac{1-A_\star\sqrt{2/3}}{1-A_\star/\sqrt{2}},
\end{equation}
which depends on the depth of the stellar line. The maximum modeled absorption, computed by a direct convolution, can thus be overestimated by a factor of up to 1.6 ultimately biasing derived parameters. This also explains why the strength of the convolution bias is reduced with shallower stellar lines (Fig.~\ref{fig:stellar_spectra}), as illustrated in Sect.~\ref{sec:prediction}.

\subsection{WASP-121 spectra for broadband features}

\begin{center}
    \includegraphics[width=\columnwidth]{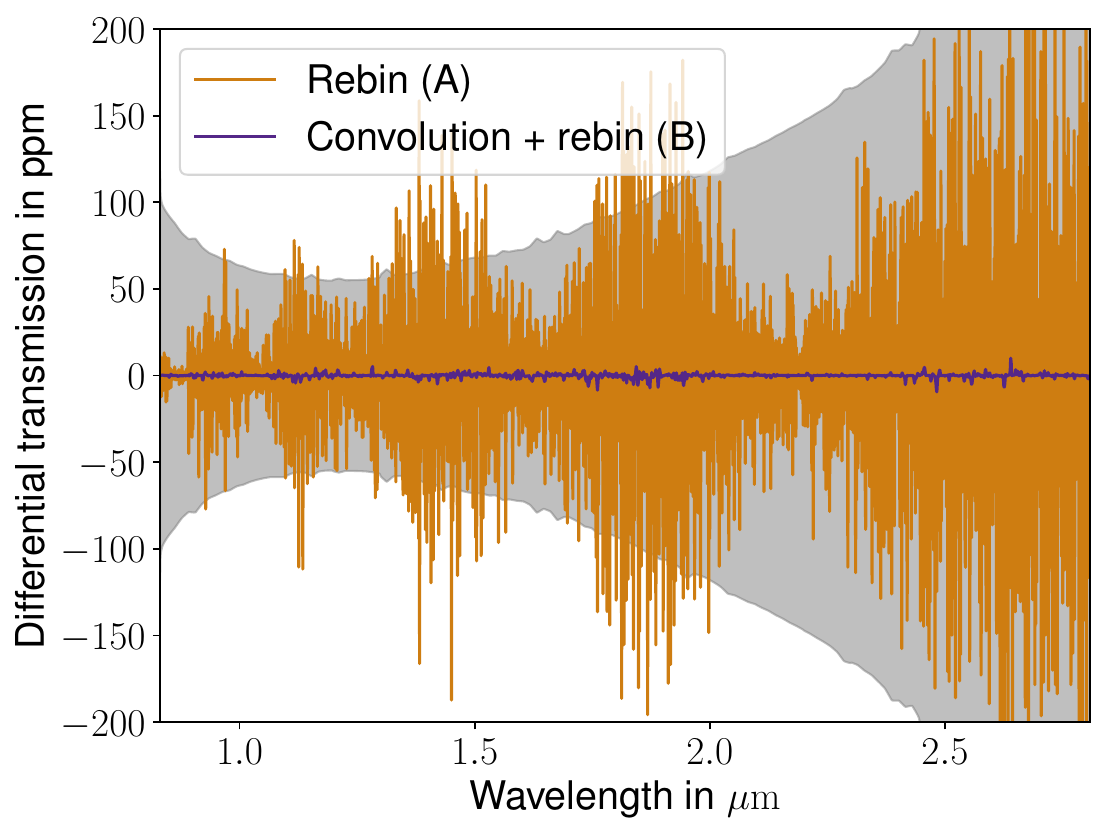}
    \centering
    \captionof{figure}{Same figure as Fig.~\ref{fig:convo_molec} but using WASP-121 stellar spectrum in Eq.~\ref{eq:abs_sp} to compute the absorption spectra.}
    \label{fig:convo_molecW121}
\end{center}

\subsection{Stellar spectra}

\begin{center}
    \includegraphics[width=\columnwidth]{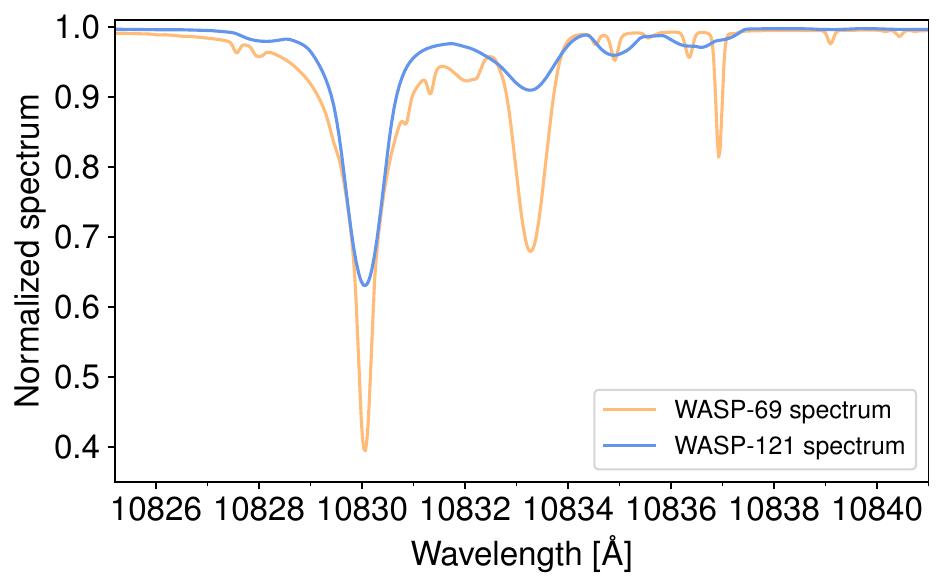}
    \centering
    \captionof{figure}{Normalized stellar spectra for the two stars considered in Sect.~\ref{sec:prediction}. The two stellar spectra are generated using \texttt{Turbospectrum} \citep{plez_turbospectrum_2012}}
    \label{fig:stellar_spectra}
\end{center}

\section{Impact of the FWHM}
\label{apdx:FWHM_prediction}

\begin{figure}
    \includegraphics[width=\columnwidth]{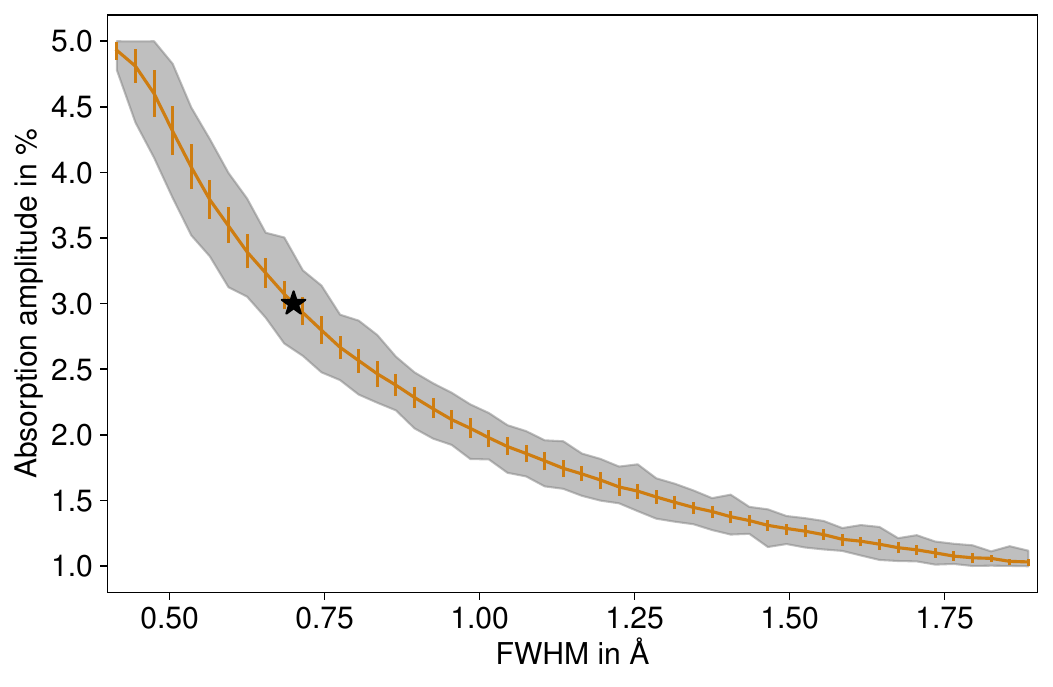}
    \centering
    \caption{Relation between the predicted absorption amplitude and FWHM at HR from a synthetic NIRSpec observation (orange line in Fig.~\ref{fig:convo_gauss}). The black star denotes the initial parameters used to generate the NIRPSEC observation. The shaded area shows the full envelop of retrieved parameters, whereas error-bars indicate $1-\sigma$ dispersion. }
    \label{fig:FWHM}
\end{figure}

In this appendix, we illustrate the dependency of the predicted He\,\textsc{i} amplitude on the FWHM. As mentioned in Sect.~\ref{sec:prediction}, various processes can increase the FWHM of the He\,\textsc{i} absorption. Typically, for sub-Neptunes or extended outflows, the FWHM might deviate significantly from the commonly assumed $\sim0.7$\,\AA. We ran a retrieval similar to that performed in Sect.~\ref{sec:prediction}, but relaxing the assumption on the FWHM. As a reminder, we generated a mock observation at NIRSpec resolution using the WASP-69 stellar spectrum and Eq.~\ref{eq:abs_sp}. The mock absorption assumes a peak absorption of 3\% and a FWHM of 0.7\,\AA. In contrast to Sect.~\ref{sec:prediction}, we fit the mock observation using the stellar spectrum as described in Eq.~\ref{eq:abs_sp}, avoiding the convolution bias. We show in Fig.~\ref{fig:FWHM} the relation between the predicted FWHM and amplitude of HR absorption from LR observations. In particular, this illustrates that the predicted HR signal amplitude can vary by up to a factor of 5, depending on the assumed FWHM. While this should not strongly impact the threshold for detection (as argued in Sect.~\ref{sec:detect_threshold}) this can significantly affect the in-depth characterization, typically preventing the study of dynamics.

\begin{figure*}[!h]
    \includegraphics[width=2\columnwidth]{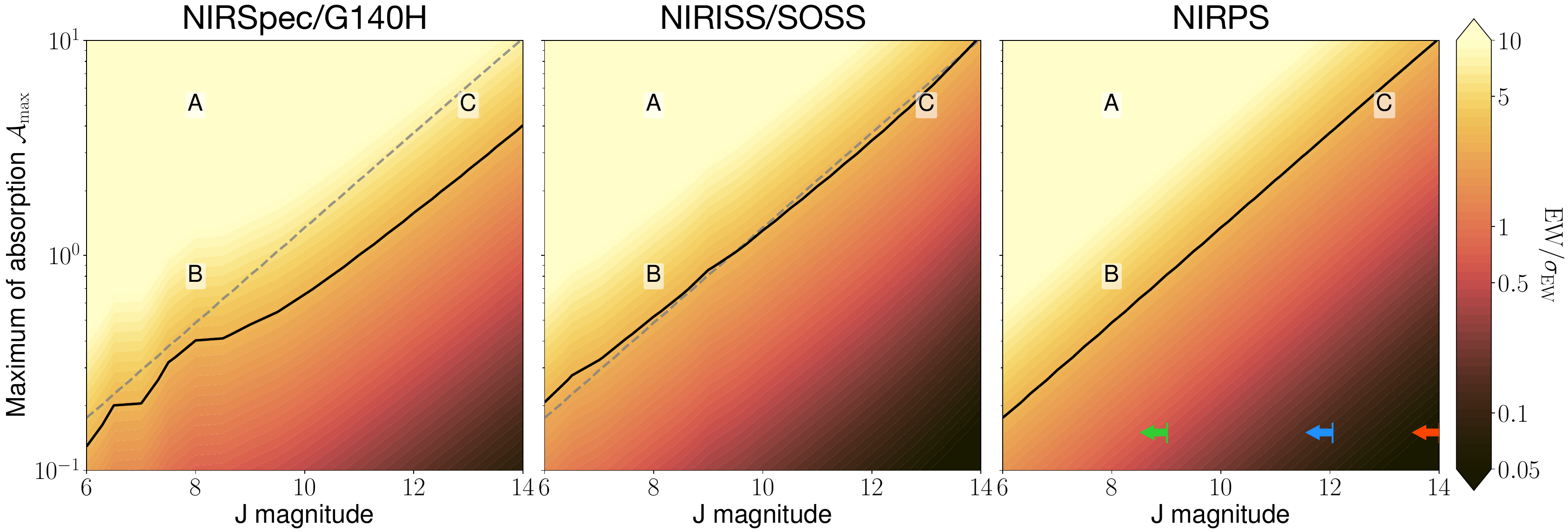}
    \centering
    \caption{Same figure as Fig.~\ref{fig:sensitivity_maps} but using the typical sub-Neptune GJ3090\,b as reference system. }
    \label{fig:sensitivity_maps_subNept}
\end{figure*}

Finally, to further illustrate how the choice of the FWHM impacts our results, we reproduced Fig.~\ref{fig:sensitivity_maps} using the sub-Neptune GJ3090\,b as the reference system in Fig.~\ref{fig:sensitivity_maps_subNept}. The FWHM of the \texttt{p-winds} absorption spectrum is $\sim 1.0$\,\AA, which is broader than that of WASP-69\,b. This illustrates that our detection limits should not be taken directly but that the comparison between instruments remains valid for a broad range of planets.

\section{Outflows structure}
\label{apdx:hr_prediction}

\begin{figure*}
    \includegraphics[width=1.6\columnwidth]{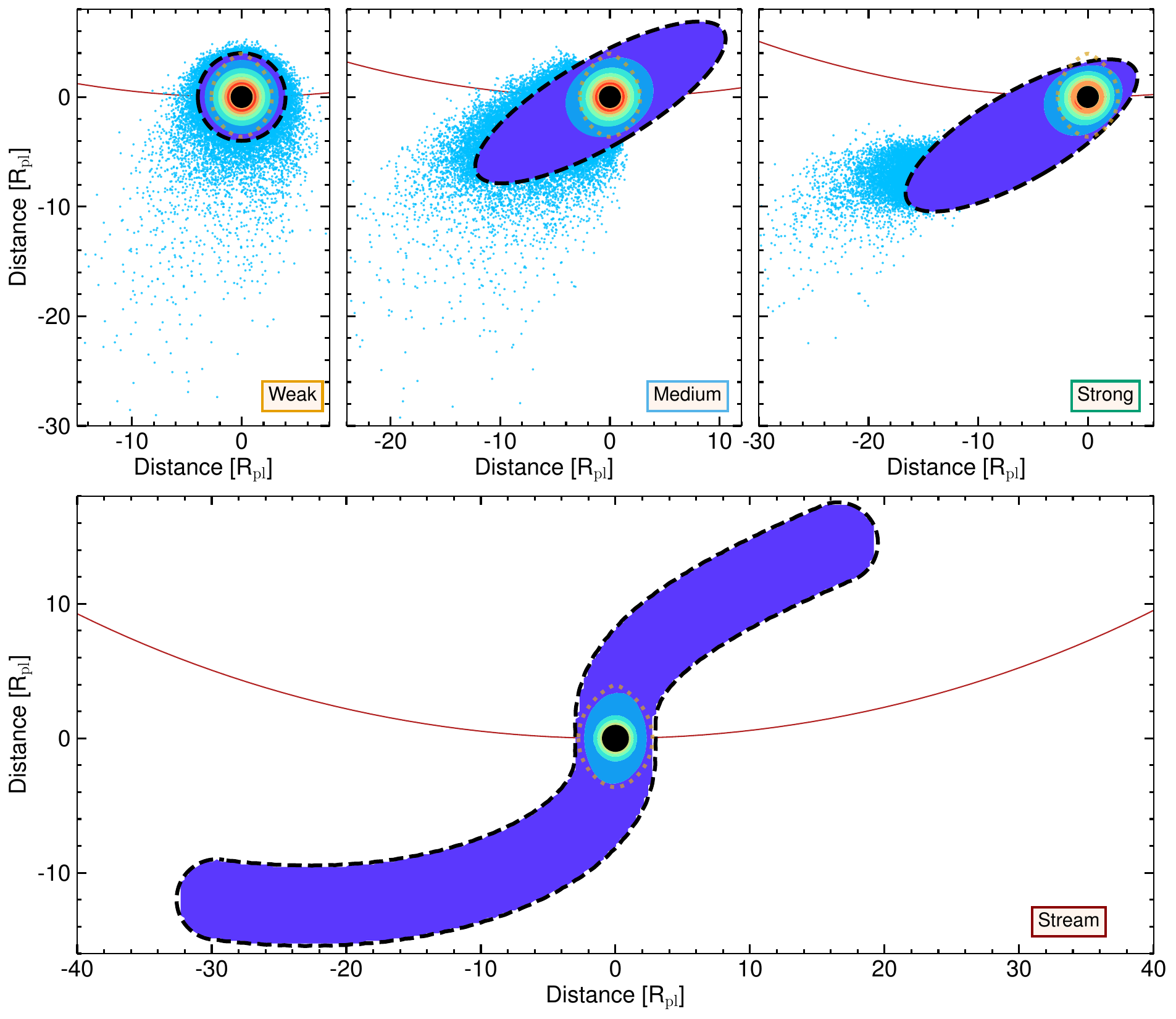}
    \centering
    \caption{Top view of the 4 outflow scenarios considered in Sect.~\ref{sec:extended_outflow}. The sky-blue dots represents particle of the exosphere, while the continuous color-map scales with the surface density of He\,\textsc{i}. In the top 3 cases the same thermospheric profile is used. In each panel, the red line shows the planetary orbit, the black dot denotes the opaque body, the dashed black line marks the exobase, and the orange dashed line is the Roche lobe. Note that the color code used for each outflow cases is the same throughout the paper.}
    \label{fig:shape}
\end{figure*}

\begin{figure*}
    \includegraphics[width=1.5\columnwidth]{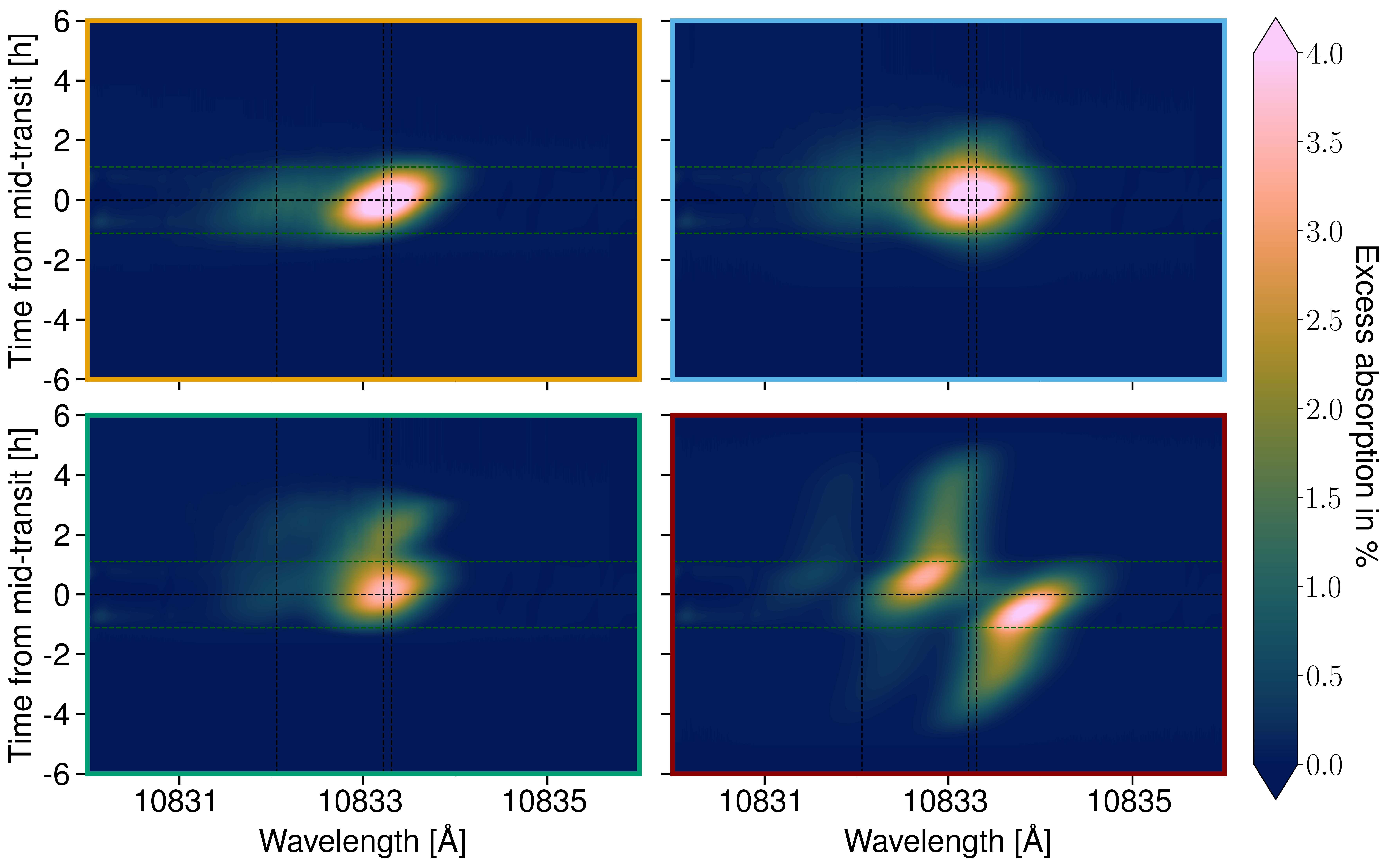}
    \centering
    \caption{Excess absorption maps of the 4 outflow scenarios considered in Sect.~\ref{sec:extended_outflow}. The color code used for the frames is the same as in Fig.~\ref{fig:shape}.}
    \label{fig:maps}
\end{figure*}

In this appendix, we present the four scenarios simulated in Sect.~\ref{sec:extended_outflow}. Fig.~\ref{fig:shape} shows top views of the thermosphere for each scenario, together with the exospheric particles. In Fig.~\ref{fig:maps}, we display the excess absorption maps at the native resolution of the simulation. These maps are then resampled temporally and convolved with the different instrumental profiles to generate mock observations.

\section{Absorption spectra in different S/N regimes}

\begin{figure*}
    \includegraphics[width=2\columnwidth]{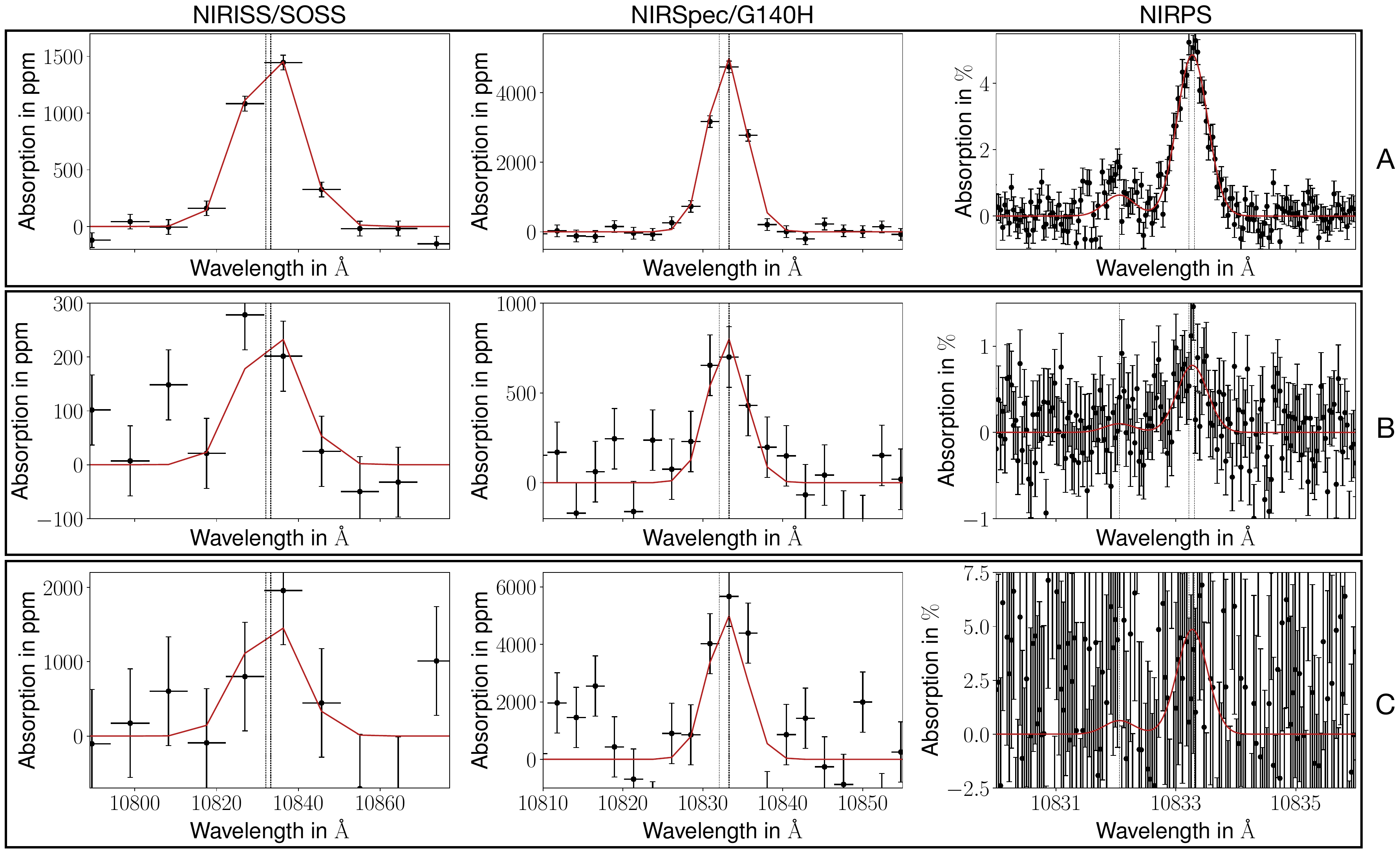}
    \centering
    \caption{Simulated average absorption spectra in the planetary rest-frame as observed with JWST/NIRISS ($\mathcal{R} \sim 650$, left), JWST/NIRSpec ($\mathcal{R} \sim 2\,050$, middle), and NIRPS ($\mathcal{R} \sim 75\,000$, right). The case A (upper panel) corresponds to high absorption amplitude and S/N, case B (middle panel) to low absorption amplitude and high S/N, and case C to high absorption amplitude and low S/N (see also Figs.~\ref{fig:sensitivity_maps}~\&~\ref{fig:sensitivity_metric}).}
    \label{fig:abs_features}
\end{figure*}

\end{appendix}

\end{document}